\documentclass[12pt,a4paper]{article}

\usepackage{arxiv}

\usepackage[utf8]{inputenc} 
\usepackage[T1]{fontenc}    
\usepackage{hyperref}       
\usepackage{url}            
\usepackage{booktabs}       
\usepackage{amsfonts}       
\usepackage{nicefrac}       
\usepackage{microtype}      
\usepackage{graphicx}
\usepackage{natbib}
\usepackage{doi}

\usepackage{newtxmath}
\usepackage{subeqnarray}

\title{Forecasting the evolution of three-dimensional  turbulent recirculating flows from sparse sensor data}

\date{} 					

\author{{George Papadakis}\thanks{corresponding author} \\
	Department of Aeronautics \\
	Imperial College London \\
	London SW7 2AZ, U.K. \\
	\texttt{g.papadakis@ic.ac.uk} \\
	\And
	{Shengqi Lu} \\
	Department of Aeronautics \\
	Imperial College London \\
	London SW7 2AZ, U.K.  \\
	\texttt{s.lu19@ic.ac.uk} \\
}

\renewcommand{\headeright}{}
\renewcommand{\undertitle}{}

\hypersetup{
	pdftitle={Forecasting the evolution of three-dimensional  turbulent recirculating flows from sparse sensor data},
	pdfsubject={},
	pdfauthor={George Papadakis, Shengqi Lu},
	pdfkeywords={First keyword, Second keyword, More},
}

\begin{document}
	\maketitle
	
\begin{abstract}
A data-driven algorithm is proposed that employs sparse data from velocity and/or scalar sensors to forecast the future evolution of three dimensional turbulent flows. The algorithm combines time-delayed embedding together with Koopman theory and linear optimal estimation theory. It consists of 3 steps;  dimensionality reduction (currently POD), construction of a linear dynamical system for current and future POD coefficients and system closure using sparse sensor measurements. In essence, the algorithm establishes a mapping from current sparse data to the future state of the dominant structures of the flow over a specified time window. The method is scalable (i.e.\ applicable to very large systems), physically interpretable, and provides sequential forecasting on a sliding time window of prespecified length. It is applied to the turbulent recirculating flow over a surface-mounted cube (with more than $10^8$ degrees of freedom) and is able to forecast accurately the future evolution of the most dominant structures over a time window at least two orders of magnitude larger that the (estimated) Lyapunov time scale of the flow. Most importantly, increasing the size of the forecasting window only slightly reduces the accuracy of the estimated future states. Extensions of the method to include convolutional neural networks for more efficient dimensionality reduction and moving sensors are also discussed.
\end{abstract}

		
	\section{Introduction}
Flow reconstruction from sparse data has received a lot of attention and the literature on the subject is vast, 
see for example \cite{Brunton_Noack_2015, Sipp_Schmid_2016, Callaham_et_al_2019, Karniadakis_et_al_2021} and references therein. On the other hand, forecasting the future evolution of the flow has received considerably less attention in fluid dynamics research, especially turbulence. The ability to estimate the future state from current data, also known as forecasting  \citep{Box_Jenkins_Reinsel_Ljung_2015} has many applications. Examples include weather forecasting, urban safety (prediction of toxic pollutant trajectory than can guide evacuation measures),  prediction of extreme events (such as heat waves) ahead of time etc.

The forecasting time window of turbulent flows is limited by a fundamental physical constraint, which arises from the fact that such flows are chaotic and thus have extreme sensitivity to initial conditions. This is popularly known as the 'butterfly effect' \citep{Lorenz1963DeterministicFlows} and can be understood as follows. Consider two turbulent flows with initial conditions that are infinitesimally close. Their trajectories in phase space will diverge at a rate determined by the maximum Lyapunov exponent, $\lambda_1$, that is $|\delta x (t)| \sim e^{\lambda_1 t}$, where $|\delta x (t)|$ is the norm of the distance between two points in phase space. The corresponding time scale is $1/\lambda_1$. 

There are theoretical arguments ~\citep{Ruelle_1979, Crisanti_et_al_1993, Ge_Rolland_Vassilicos_2023} and computational evidence~\citep{Mohan_et_al_2017,Hassanaly_2019} that in turbulent flows $\lambda_1$ scales with the Kolmogorov time scale, $\tau$ (with a correction factor that depends on Reynolds number). Therefore the future evolution of turbulent flows can only be forecast over a time window that is only several times larger than $\tau$, say $(6-10) \tau$. This has been confirmed repeatedly irrespective of the method used for forecasting, see \cite{Eivazi_et_al_2021, Vlachas_et_al_2020,Pathak_et_al_2017,Khodkar2021,Dubois_et_al_2020} for a very small sample of extensive literature on the subject. Therefore for high Reynolds number flows this time window is very short to be useful in practice. For other applications however, for example in medium-range weather forecasting, current lead times are 10 days, but can reach up to 15 days if the uncertainty in the initial conditions is reduced by an order of magnitude \citep{Allen_et_al_2025, Zhang_Sun_et_al_2019}. This window is sufficiently large to bring about enormous socioeconomic benefits, for example protecting lives, property etc.

Is it possible to enlarge the forecasting window for turbulent flows and get estimations that are useful in engineering practice? The aforementioned scaling of $\lambda_1$ with $\tau$ indicates that the former is determined by the smallest scales of the flow. However a turbulent flow consists of a wide range of spatio-temporal structures, and it is well known that large scale structures are slower and more organised, \cite{pope_2000}.  This opens the possibility that they can be more amenable to forecasting. These structures are important because they determine momentum transfer (and therefore forces) and  scalar dispersion. The ability therefore to forecast their future evolution can bring many practical benefits. Most of existing work on forecasting to date has focused on low-dimensional dynamical systems (such the Lorenz 63/96 systems, the Kuramoto-Sivashinsky equation, the 9-equation model of \cite{Moehlis_et_al_2004}) or two-dimensional flows (such as flow in a 2D lid-driven cavity). These dynamical systems however do not exhibit the large separation of length and time scales that is necessary to answer the question at the beginning of this paragraph. To fill this gap, the present paper considers the three-dimensional turbulent recirculating flow around a surface-mounted prism. As will be seen later, this flow contains a wide range of scales, such as large organised structures and a clear inertial regime. 

Naive application of existing dimensionality reduction methods, such as Dynamic Mode Decomposition (DMD), for forecasting will fail. DMD constructs a linear model of the form $\boldsymbol{x}[k+1]=A \boldsymbol{x}[k]$, see \cite{Schmid2010JFM}, where $\boldsymbol{x}[k]$ is the state vector at time instant $k$. Recursive application of this relation results in exponentially increasing or decreasing state values (depending on the eigenvalues of $A$). In the case of exponentially increasing values, the growth rate is not related to $\lambda_1$. The same problem appears for higher order DMD, which can be put in the same linear form as the standard DMD, but now $\boldsymbol{x}[k]$ contains time-delayed variables, \cite{LeClainche_Vega_2017}. 

More advanced approaches are therefore necessary. In particular, the decomposition of chaotic dynamics into a linear model with forcing is a very attractive approach because of the availability of a large number of tools for linear systems. 
Recently \cite{Chu_Schmidt_2025} derived a linear, time-invariant model for the coefficients of Spectral Proper Orthogonal Decomposition (SPOD) modes. The authors retained the forcing term and included its dynamics in the model formulation. The presence of stochastic forcing term results in a probabilistic representation of the future evolution of dominant structures. \cite{Brunton_et_al_2017} combined time-delay embedding and Koopman theory to derive a linear model in the leading time-delay coordinates which is forced by low-energy variables. The model states are obtained from the right singular vectors of the Hankel matrix which is assembled by stacking time-delayed values of observables column by column (each column is advanced one time unit ahead of the previous column). The forcing was active only in the regions of the chaotic attractor with strong non-linearity, and its statistics were found to be non-Gaussian. The authors call this the Hankel alternative view of Koopman (HAVOK).  In \cite{Khodkar2021} the forcing is found from vector-valued observables in a physics-informed way or a purely data-driven fashion, depending on whether any knowledge of governing dynamics was available or not. In \cite{Dylewsky_et_al_2022} the forcing is obtained in a two-step process in a fully unsupervised manner, again using the measurement data only.

The usefulness of time-delayed embeddings of even a single  observable for the analysis of chaotic systems was recognized in \cite{Takens_1981}. The combination with Koopman theory in particular has opened new directions for the representation of a chaotic system as a forced linear model and has spawned important theoretical work \citep{Kamb_et_al_2020, Arbabi2017, Pan_Duraisamy_2019, Pan_Duraisamy_2020,Giannakis_2019, Das_Giannakis_2019}. The connection between the left-singular vectors of the time-delayed Hankel matrix with the space-time POD, space-only POD and SPOD modes was established in \cite{Frame_Towne_2023}. 

The contribution of the present paper is to add one more piece to the aforementioned time-delayed embedding/Koopman framework. More specifically, we combine this framework with linear optimal estimation theory  \citep{Kailath_Hassibi_Sayed_2000} that allows us to derive a mapping from sparse measurements at the current time instant to the velocity field at future time instants. This approach is useful because it circumvents the need to estimate the forcing term of the linear system. \cite{Schmidt_2025} also proposed a method that does not require an estimation of the forcing term. This was achieved by leveraging the correlation between hindcast and forecast datasets with the aid of extended POD. To forecast the future evolution of the flow, data are required for the full flow, while in our method, once the mapping has been established, only data from a few sensors are required. A non-linear mapping that employs machine learning techniques (instead of Koopman theory) was proposed recently for weather forecasting by \cite{Allen_et_al_2025}.  

Our approach is applied to flow around a surface-mounted cube, as already mentioned. Scalar is released from a source placed upstream of the cube. We derive the mapping from current to future states using not only velocity data but also scalar data; scalar sensors are usually cheaper to acquire. The proposed estimator is scalable, physically interpretable, and provides sequential forecasting on a rolling time window as data are coming in. Of particular interest is the quality  of prediction and how it varies with the size of the time window.

The paper is structured as follows. Sections \ref{sec:forecast_velocity} abd \ref{subsec:EstimateFromSca} describe the forecasting methodology from sparse velocity and scalar data respectively. Results are presented and discussed in \ref{sec:flow_surf_mounted_cube}. Section \ref{sec:conclusions} summarizes the main findings of the paper and outlines some future research directions.

\section{Flow forecasting from sparse velocity measurements}{\label{sec:forecast_velocity}}

In the following, $u,v,w$ (interchangeably used with $u_1,u_2,u_3$) are the three velocity components in the Cartesian directions $x,y,z$ respectively. Time-averages are designated with angular brackets and fluctuations with a prime;  for  example $\langle u \rangle$, $u^\prime$ are the mean and fluctuating velocities respectively in the $x$ direction.

We first describe how to forecast the future evolution of a turbulent flow from a set of sparse velocity measurements that record current information. In the following section, we extend the idea to sparse scalar measurements. 

The method comprises three steps. 

\subsection {Step 1: Dimensionality reduction.}
To make the method applicable for three-dimensional turbulent flows, we first need to reduce the number of degrees of freedom. In this paper we use using Proper Orthogonal Decomposition (POD), \citet{Sirovich1987}. This method  provides a linear mapping between the velocity field and the POD coefficients; we exploit this linearity later in step 3. Other techniques, such as convolutional autoencoders \citep{Brunton_et_al_2020}, can be also employed. They are more efficient in terms of dimensionality reduction, but they result in a non-linear mapping between the new degrees of freedom (latent variables) and the velocity field. Our approach can still be used but with some modifications, as explained later.

To get the POD modes, the snapshot matrix $\boldsymbol{Y}(\boldsymbol{x},t_{1}:t_{K})$ for the velocity fluctuations $u^{\prime}$, $v^{\prime}$ and $w^{\prime}$ is assembled,  
\begin{equation}
	\boldsymbol{Y}(\boldsymbol{x},t_{1}:t_{K})
	=
	\left[ \begin{array}{cccc}
		u^{\prime}(\boldsymbol{x}_{1}, t_{1})&
		u^{\prime}(\boldsymbol{x}_{1}, t_{2})&
		\ldots&
		u^{\prime}(\boldsymbol{x}_{1}, t_{K})\\
		\vdots & \vdots & \ldots & \vdots\\
		u^{\prime}(\boldsymbol{x}_{N}, t_{1})&
		u^{\prime}(\boldsymbol{x}_{N}, t_{2})&
		\ldots&
		u^{\prime}(\boldsymbol{x}_{N}, t_{K})\\
		v^{\prime}(\boldsymbol{x}_{1}, t_{1})&
		v^{\prime}(\boldsymbol{x}_{1}, t_{2})&
		\ldots&
		v^{\prime}(\boldsymbol{x}_{1}, t_{K})\\ 
		\vdots & \vdots & \ldots & \vdots\\    
		v^{\prime}(\boldsymbol{x}_{N}, t_{1})&
		v^{\prime}(\boldsymbol{x}_{N}, t_{2})&
		\ldots&
		v^{\prime}(\boldsymbol{x}_{N, t_{K}})\\
		w^{\prime}(\boldsymbol{x}_{1}, t_{1})&
		w^{\prime}(\boldsymbol{x}_{1}, t_{2})&
		\ldots&
		w^{\prime}(\boldsymbol{x}_{1}, t_{K})\\ 
		\vdots & \vdots & \ldots & \vdots\\    
		w^{\prime}(\boldsymbol{x}_{N}, t_{1})&
		w^{\prime}(\boldsymbol{x}_{N}, t_{2})&
		\ldots&
		w^{\prime}(\boldsymbol{x}_{N, t_{K}})    
	\end{array}  \right] ,
	\label{eq: velocity_snapshot_matrix}
\end{equation}
\noindent where $\boldsymbol{Y} \in \mathbb{R}^{3N \times K}$, $\boldsymbol{x}_{i}=
\left[x_{i}, y_{i}, z_{i} \right] (i=1,2,...N)$ is the location vector for the $i$-th spatial location, $N$ is the number of cells, and $K$ is the total number of snapshots. The spacing between successive snapshots is $\Delta t$. Singular value decomposition is performed on the weighted matrix $\mathcal{V}^{1/2} \boldsymbol{Y}$,
\begin{equation}
	\mathcal{V}^{1/2} \boldsymbol{Y}=
	\boldsymbol{U}_{Y} \boldsymbol{ \mathit{\Sigma} }_{Y}
	\boldsymbol{V}^{\top}_{Y},
	\label{eq: velocity_SVD_eqn}
\end{equation}

\noindent where $\mathcal{V}=diag \left(V_1, V_2 \dots V_N, V_1, V_2, \dots, V_N, V_1, V_2, \dots, V_N \right)$ is a diagonal matrix with the cell volumes $V_i$ in the main diagonal, $\boldsymbol{U}_{Y} \in \mathbb{R}^{3N \times K}$
contains the left singular vectors, $\boldsymbol{ \mathit{\Sigma} }_{Y} \in \mathbb{R}^{K \times K}$
is a diagonal matrix that stores $K$ singular values, and $\boldsymbol{V}_{Y} \in \mathbb{R}^{K \times K}$
contains the right singular vectors. The scaled POD eigenmodes $U_{Y, k}(\boldsymbol{x})$ are extracted from the columns of $\boldsymbol{U}_{Y}$ from, 
\begin{equation}
	{U}_{Y} (\boldsymbol{x})
	=
	\mathcal{V}^{-1/2}
	\boldsymbol{U}_{Y}. 
\end{equation}

The singular values $\sigma_{Y, k}$ are ranked in descending order along the diagonal of matrix $\boldsymbol{ \mathit{\Sigma} }_{Y}$. The energy content of each mode is computed using,  
\begin{equation}
	\lambda_{Y, k}=\frac{ \sigma^{2}_{Y, k} }{K}, \quad (k=1\dots K)
\end{equation}
and the time coefficients from,
\begin{equation}
	\boldsymbol{a} (t)=
	\boldsymbol{ \mathit{Y} }^{\top}
	\mathcal{V}^{1/2}
	\boldsymbol{U}_{Y}.
	\label{eq:vel_time_coeff}
\end{equation}

The fluctuating velocity field can be written as,
\begin{equation} 
	u^{\prime}_{i}(x,y,z,t)
	=
	\sum^{ K }_{k=1}
	a_{k}(t) {U}^{(i)}_{Y, k}(x,y,z)
	\approx
	\sum^{m_{u}}_{k=1}
	a_{k}(t) {U}^{(i)}_{Y, k}(x,y,z), \: (i=1, 2, 3),
	\label{eq:uprime_sum_modes}
\end{equation}

\noindent where $m_{u}$ is the number of retained POD modes, $a_{k}(t)$ is the time coefficient of the $k$-th POD mode, and ${U}^{(i)}_{Y, k}$ is the $k$-th POD eigenvector of the $i$-th velocity component. This expression can be written in matrix form as,

\begin{equation}
	\underbrace{\left[
		\begin{array}{c}
			u_1^{\prime} \\    
			u_2^{\prime}\\   
			u_3^{\prime}\end{array}  \right]}
	_{\equiv \boldsymbol{u}^{\prime}} =
	\underbrace{\left[
		\begin{array}{cccc}
			{U}^{(1)}_{Y, 1} &  {U}^{(1)}_{Y, 2} & \dots &  {U}^{(1)}_{Y, m_u} \\ 
			{U}^{(2)}_{Y, 1} &  {U}^{(2)}_{Y, 2} & \dots &  {U}^{(2)}_{Y, m_u}    \\   
			{U}^{(3)}_{Y, 1} &  {U}^{(3)}_{Y, 2} & \dots &  {U}^{(3)}_{Y, m_u} 
		\end{array}  \right]}_{\equiv \boldsymbol{U}_{Y}}
	\underbrace{\left[
		\begin{array}{c}
			a_1^{\prime} \\    
			a_2^{\prime}\\  
			\vdots \\
			a_{m_u}^{\prime}\end{array}  \right]}_{\equiv \boldsymbol{a}}
	\label{eq:velo_pod_coef}
\end{equation}
or more compactly
\begin{equation}
	\boldsymbol{u}^{\prime}(x,y,z,t)=\boldsymbol{U}_{Y}(x,y,z) \boldsymbol{a}(t).
	\label{eq:velo_pod_coef_2}
\end{equation}

\subsection{Step 2: Construction of a dynamical system for current and future POD coefficients.}
The next step is the construction of a model for the dynamic evolution of POD coefficients. This model should be able to forecast the future development from current information. To this end, we assemble the time-delayed Hankel matrix $\boldsymbol{H}$ that consists of the POD coefficients of the retained $m_{u}$ velocity modes,
\begin{equation}
	\boldsymbol{H}
	=
	\left[
	\begin{array}{cccc}
		\boldsymbol{a}(t_{1})& \boldsymbol{a}(t_{2})& \ldots& \boldsymbol{a}(t_{p})\\    
		\boldsymbol{a}(t_{2})& \boldsymbol{a}(t_{3})& \ldots&
		\boldsymbol{a}(t_{p+1})\\   
		\vdots& \vdots& \ddots& \vdots\\
		\boldsymbol{a}(t_{q})& \boldsymbol{a}(t_{q+1})& \ldots&
		\boldsymbol{a}(t_{K_{train}})
	\end{array}  \right] ,
	\label{eq:H_detail}
\end{equation}
\noindent where $ \boldsymbol{a}(t_{j})=\left [a_1(t_j)\dots a_{m_u}(t_j) \right]^\top$ and we use $q$ vectors in each column. The number of columns is  $p=K_{train}-q+1$ where $K_{train}$ is the number of snapshots for the training data set, and $\boldsymbol{H} \in \mathbb{R}^{\left( m_{u} \times q \right) \times p}$. 
Performing SVD on $\boldsymbol{H}$ we obtain,
\begin{equation}
	\boldsymbol{H}
	\approx
	\boldsymbol{U}_{H} \boldsymbol{\Sigma}_{H}
	\boldsymbol{V}^{\top}_{H},
\end{equation}

\noindent where $\boldsymbol{U}_{H} \in \mathbb{R}^{\left( m_{u} \times q \right) \times r}$, $\boldsymbol{\Sigma}_{H} \in \mathbb{R}^{r \times r}$, $\boldsymbol{V}_{H} \in \mathbb{R}^{p \times r}$ and $r$ is the number of retained singular values. The matrix of the left singular vectors $\boldsymbol{U}_{H}$ can be  explicitly written as,
\begin{equation}
	\boldsymbol{U}_{H}
	=
	\left[ \begin{array}{cccc}
		\boldsymbol{U}^{(u,v,w)}_{H, 1}(t_{1})& \boldsymbol{U}^{(u,v,w)}_{H, 2}(t_{1})& \ldots&
		\boldsymbol{U}^{(u,v,w)}_{H, r}(t_{1})\\
		\boldsymbol{U}^{(u,v,w)}_{H, 1}(t_{2})& \boldsymbol{U}^{(u,v,w)}_{H, 2}(t_{2})& \ldots&
		\boldsymbol{U}^{(u,v,w)}_{H, r}(t_{2})\\    
		\vdots& \vdots& \ddots& \vdots\\
		\boldsymbol{U}^{(u,v,w)}_{H, 1}(t_{q})& \boldsymbol{U}^{(u,v,w)}_{H, 2}(t_{q})& \ldots&
		\boldsymbol{U}^{(u,v,w)}_{H, r}(t_{q})  
	\end{array}  \right] ,
	\label{eq:U_H_detail}
\end{equation}
\noindent where each column $\boldsymbol{U}^{(u,v,w)}_{H,i}(t_{1} \colon t_{q}) \in \mathbb{R}^{m_{u} \times q }$ is the $i$-th time-delayed singular mode of the Hankel matrix.

The diagonal matrix of singular values $\boldsymbol{\Sigma}_{H}$ is, 
\begin{equation}
	\boldsymbol{\Sigma}_{H}
	=
	\left[ \begin{array}{cccc}
		\sigma_{H, 1}& 0& \ldots& 0\\
		0& \sigma_{H, 2}& \ldots& 0\\
		\vdots& \vdots& \ddots& \vdots\\
		0& 0& \ldots& \sigma_{H, r}
	\end{array}  \right] ,
\end{equation}

and the matrix of the right singular vectors $\boldsymbol{V}_{H}$ can be explicitly written as, 
\begin{equation}
	\boldsymbol{V}_{H}
	=
	\left[ \begin{array}{cccc}
		v_{H,1}(t_{1})& v_{H,1}(t_{2})& \ldots& v_{H,1}(t_{p})\\
		v_{H,2}(t_{1})& v_{H,2}(t_{2})& \ldots& v_{H,2}(t_{p})\\
		\vdots& \vdots& \ddots& \vdots\\
		v_{H,r}(t_{1})& v_{H,r}(t_{2})& \ldots& v_{H,r}(t_{p})
	\end{array}  \right] .
\end{equation}

We now define the vector $\boldsymbol{v}_{H}(t_j)=[v_{H, 1}(t_j), v_{H, 2}(t_j), \ldots, v_{H, r}(t_j)]^{\top} \in \mathbb{R}^{r}$ $(j=1 \dots p)$ extracted from the columns of matrix $\boldsymbol{V}$. We consider $\boldsymbol{v}_{H}$ as the state variable of the following discrete in time, forced, linear dynamical system, 
\begin{equation}
	\boldsymbol{v}_{H}[k+1]
	=
	\boldsymbol{A} \boldsymbol{v}_{H}[k]+  \boldsymbol{w}_{2}[k],
	\label{eq:noForcingTerm}
\end{equation}
\noindent where $\boldsymbol{v}_{H}[k]=\boldsymbol{v}_{H}[t_k]$, $k=1 \dots p-1$   and $\boldsymbol{A} \in \mathbb{R}^{r \times r}$.

To obtain matrix $\boldsymbol{A}$ we write \eqref{eq:noForcingTerm} for $k=1 \dots p-1$.
\begin{subeqnarray}
	\boldsymbol{v}_{H}[2] 
	& \approx &
	\boldsymbol{A}\boldsymbol{v}_{H}[1],\\
	\boldsymbol{v}_{H}[3] 
	& \approx &
	\boldsymbol{A}\boldsymbol{v}_{H}[2],\\
	& \vdots &\nonumber\\
	\boldsymbol{v}_{H}[p] 
	& \approx & 
	\boldsymbol{A}\boldsymbol{v}_{H}[p-1].
	\label{eq:discreteEqnNoForcing}
\end{subeqnarray}

In matrix form, this becomes, 
\begin{equation}
	\boldsymbol{V}^{\prime}
	\approx
	\boldsymbol{A} \boldsymbol{V},
	\label{eq:matNoForcing}
\end{equation}
\noindent where $\boldsymbol{V}^{\prime}, \boldsymbol{V} \in \mathbb{R}^{r \times (p-1)}$.
\begin{subeqnarray}
	\boldsymbol{V}^{\prime}
	& = &
	\left[ \begin{array}{cccc}
		\boldsymbol{v}_{H}[2]& 
		\boldsymbol{v}_{H}[3]& 
		\ldots& 
		\boldsymbol{v}_{H}[p]
	\end{array}  \right]^{\top},\\
	\boldsymbol{V}
	& = &
	\left[ \begin{array}{cccc}
		\boldsymbol{v}_{H}[1]&
		\boldsymbol{v}_{H}[2]&
		\ldots&
		\boldsymbol{v}_{H}[p-1]
	\end{array}  \right]^{\top}.
\end{subeqnarray}

Then the system matrix $\boldsymbol{A}$ can be calculated from, 
\begin{equation}
	\boldsymbol{V}^{\prime}
	=
	\boldsymbol{A} {\boldsymbol{V}}
	\Rightarrow
	\boldsymbol{V}^{\prime} {\boldsymbol{V}}^{\top}
	=
	\boldsymbol{A} {\boldsymbol{V}} {\boldsymbol{V}}^{\top}
	\Rightarrow
	\boldsymbol{A}
	=
	\boldsymbol{V}^{\prime} {\boldsymbol{V}}^{\top}
	\big(
	{\boldsymbol{V}} {\boldsymbol{V}}^{\top}
	\big)^{-1}.
\end{equation}

This formulation applies the DMD algorithm directly to $ \boldsymbol{v}_{H}$. 

Once $\boldsymbol{A}$ is known, the forcing $\boldsymbol{w}_{2}[k]$ can be easily computed from the training data set, $\boldsymbol{w}_{2}[k]=\boldsymbol{v}_{H}[k+1]-\boldsymbol{A} \boldsymbol{v}_{H}[k]$. The covariance matrix of the forcing $\boldsymbol{Q} \in \mathbb{R}^{r \times r}$ can be obtained from, 
\begin{equation}
	\boldsymbol{Q}
	=
	\mathbb{E}
	\big( \boldsymbol{w}_{2}\boldsymbol{w}^{\top}_{2} \big)
	=
	\frac{1}{p-1}
	\sum^{k=p-1}_{k=1} 
	\boldsymbol{w}_{2}[k] 
	\boldsymbol{w}^{\top}_{2}[k].
\end{equation}

It is important to notice that using $\boldsymbol{v}_{H}[k]$ we can obtain the current (at instant $k$) and forecast the future (at instants $k+1 \dots k+q$) time coefficients as follows, 
\begin{eqnarray}
	\left[ 
	\begin{array}{c}
		\boldsymbol{a}[k] \\
		\boldsymbol{a}[k+1] \\
		\vdots \\
		\boldsymbol{a} [k+q] \\
	\end{array} 
	\right]
	& = &  \boldsymbol{U}_{H} \Sigma_{H} 
	\boldsymbol{v}_{H}[k].
	\label{eq:PredKal_1}
\end{eqnarray}
In particular $\boldsymbol{a}[k]$ can be obtained from,
\begin{equation}
	\boldsymbol{a}[k]
	=
	\underbrace{\boldsymbol{U}_{H}(t_{1}) \Sigma_{H}}
	_{\equiv \boldsymbol{C}}
	\boldsymbol{v}_{H}[k]
\end{equation}
\noindent where $\boldsymbol{C} \in \mathbb{R}^{m_{u} \times r}$ and  
$\boldsymbol{U}_{H}(t_{1})=
\left[ \begin{array}{cccc}
	\boldsymbol{U}^{(u,v,w)}_{H, 1}(t_{1})&
	\boldsymbol{U}^{(u,v,w)}_{H, 2}(t_{1})&
	\ldots&
	\boldsymbol{U}^{(u,v,w)}_{H, r}(t_{1})
\end{array}  \right] \in \mathbb{R}^{m_{u} \times r}$ is the top row of matrix $\boldsymbol{U}_{H}$, see \eqref{eq:U_H_detail}. 

The previous two expressions demonstrate the fundamental importance of $\boldsymbol{U}_{H}$. This matrix has a very clear physical interpretation. It encapsulates patterns (or modes) from the past $q$ time instants that can be exploited to predict the future evolution from current information. These modes are arranged column by column in terms of importance, as quantified by the singular values $\sigma_{H,k}$. They resemble Legendre polynomials for short $q \times \Delta t$ and become sinusoidal for large $q \times \Delta t$, see \cite{Dylewsky_et_al_2022, Frame_Towne_2023}. We visualise these modes for the flow around a surface-mounted cube in sections \ref{subsec:VelOnly} and \ref{subsec:ScaOnly}.

Note that this step is independent of the dimensionality reduction method selected in step 1. If a convolutional autoencoder is employed, then a time-delayed Hankel matrix can still be assembled from  $\boldsymbol{a}(t_{j})$ that now represent the latent space variables.  

The question now is how to obtain $\boldsymbol{v}_{H}[k]$ since the forcing term $\boldsymbol{w}_{2}[k]$ in \eqref{eq:noForcingTerm} is not known. We can get a closure if measurements at some sparse sensor points are available.

\subsection{Step 3: Closure of the system using sensor measurements.}
Let's assume that we have a set of $l$ velocity measurements $\boldsymbol{s}[k] \in \mathbb{R}^{l}$ at a number of a sensor points. At each sensor, one or more velocity components are recorded. We can express $\boldsymbol{s}[k]$ in terms of $\boldsymbol{a}[k]$ as follows 
\begin{equation}
	\boldsymbol{s}[k]
	=
	\boldsymbol{S} \boldsymbol{U}_{Y} \boldsymbol{a}[k]
	+
	\boldsymbol{g}[k],
	\label{eq:map_measur_coeff}
\end{equation}

\noindent where matrix $\boldsymbol{S}$ selects the rows of the POD mode matrix $\boldsymbol{U}_{Y}$ corresponding to the sensor location and the velocity component being measured (it is $0$ everywhere except for those points and components where the corresponding element is equal to $1$). Vector $\boldsymbol{g} \in \mathbb{R}^{l}$ includes measurement errors as well as errors due to POD mode truncation (because only $m_u$ modes are retained). The elements of $\boldsymbol{g}[k]$ can be obtained from the training data set, $\boldsymbol{g}[k]=\boldsymbol{s}[k]-\boldsymbol{S} \boldsymbol{U}_{Y}  \boldsymbol{a}[k]$), and its covariance $\boldsymbol{R} \in \mathbb{R}^{l \times l}$ can be easily calculated from,
\begin{equation}
	\boldsymbol{R}
	=
	\mathbb{E}
	\big( \boldsymbol{g}\boldsymbol{g}^{\top} \big)
	=
	\frac{1}{p-1}
	\sum^{k=p-1}_{k=1} 
	\boldsymbol{g}[k] 
	\boldsymbol{g}^{\top}[k].
\end{equation}

We are now ready to design a Kalman filter to estimate  $\boldsymbol{v}_{H}[k]$ from the sparse measurements $\boldsymbol{s}[k]$. The filter takes the form, 
\begin{subeqnarray}
	\hat{\boldsymbol{v}}_{H}[k+1]
	& = &
	\boldsymbol{A} \hat{\boldsymbol{v}}_{H}
	+
	\mathcal{L} \big(
	\boldsymbol{s}[k]-\hat{\boldsymbol{s}}[k]
	\big),\\
	\hat{\boldsymbol{a}}[k]
	& = &
	\boldsymbol{C} \hat{\boldsymbol{v}}_{H}[k],\\
	\hat{\boldsymbol{s}}[k]
	& = &
	\boldsymbol{S} \boldsymbol{U}_{Y} \hat{\boldsymbol{a}}[k]
	=
	\boldsymbol{S} \boldsymbol{U}_{Y} \boldsymbol{C} 
	\hat{\boldsymbol{v}}_{H}[k],
	\label{eq:filter_form}
\end{subeqnarray}

\noindent where a hat $\hat {()} $ denotes an estimated quantity and $\mathcal{L} \in \mathbb{R}^{r \times l}$ is the Kalman filter gain. The latter is obtained from the solution of the following Riccati equation, 
\begin{subequations}
	\begin{align} 
		\boldsymbol{P}
		& = 
		\boldsymbol{A} \boldsymbol{P} \boldsymbol{A}^{T}
		-
		\boldsymbol{A} \boldsymbol{P} 
		(\boldsymbol{S} \boldsymbol{U}_{Y}  \boldsymbol{C})^{T}
		(
		\boldsymbol{S} \boldsymbol{U}_{Y}  \boldsymbol{C} \boldsymbol{P}
		(\boldsymbol{S} \boldsymbol{U}_{Y}  \boldsymbol{C})^{T}
		+
		\boldsymbol{R}
		)^{-1}
		\boldsymbol{S} \boldsymbol{U}_{Y}  \boldsymbol{C} 
		\boldsymbol{P} \boldsymbol{A}^{T}
		+
		\boldsymbol{Q},\\
		\mathcal{L}
		& = 
		\boldsymbol{A} \boldsymbol{P}
		(\boldsymbol{S} \boldsymbol{U}_{Y}  \boldsymbol{C})^{T}
		(
		\boldsymbol{S} \boldsymbol{U}_{Y}  \boldsymbol{C} \boldsymbol{P}
		(\boldsymbol{S} \boldsymbol{U}_{Y}  \boldsymbol{C})^{T}
		+
		\boldsymbol{R}
		)^{-1}.
	\end{align}
	\label{eq:Riccati}
\end{subequations}
From $\hat{\boldsymbol{v}}_{H}[k]$ one can estimate the current and future POD coefficients from \eqref{eq:PredKal_1}, and of course the instantaneous velocity field from equation \eqref{eq:velo_pod_coef_2}. 

The linear mapping between the velocity and POD coefficients, equation \eqref{eq:map_measur_coeff}, has allowed us to synthesize a Kalman filter. This type of filter is suitable only for linear dynamical systems and linear mappings between the measurement and state variables. Convolutional neural networks result in non-linear mappings, but in this case one can use an extended Kalman filter or an Ensemble Kalman filter \citep{Evensen_2003}. Note also that the formulation can accommodate data from moving sensor locations (for example data from drones); this is achieved by using a time dependent selection matrix $\boldsymbol{S}[k]$ in \eqref{eq:map_measur_coeff}. In this case, the filter gain $\mathcal{L}[k]$ will be also time dependent. 

Steps 1 and 2 are performed offline using a training data set. The Kalman estimator in step 3 runs online and requires only the streaming measurement data $\boldsymbol{s}[k]$. The arriving data are then mapped to the future velocity field on a rolling time window through \eqref{eq:filter_form} to \eqref{eq:PredKal_1} and finally \eqref{eq:velo_pod_coef_2}. Depending on the number of retained modes $m_u$, it may be possible to perform such a  sequential forecasting in real time, thereby making the approach also useful to experimentalists.

\section{Flow forecasting from sparse scalar measurements}
\label{subsec:EstimateFromSca}

We follow the same approach to forecast the flow from sparse scalar measurements. First we assemble the snapshot matrix $\boldsymbol{ \mathit{Z} }(\boldsymbol{x},t_{1}:t_{K})$ of the scalar fluctuations $c^{\prime}$,
\begin{equation}
	\boldsymbol{ \mathit{Z} }(\boldsymbol{x},t_{1}:t_{K})
	=
	\left[ \begin{array}{cccc}
		c^{\prime}(\boldsymbol{x}_{1}, t_{1})&
		c^{\prime}(\boldsymbol{x}_{1}, t_{2})&
		\ldots&
		c^{\prime}(\boldsymbol{x}_{1}, t_{K})\\
		\vdots & \vdots & \ddots & \vdots\\
		c^{\prime}(\boldsymbol{x}_{N}, t_{1})&
		c^{\prime}(\boldsymbol{x}_{N}, t_{2})&
		\ldots&
		c^{\prime}(\boldsymbol{x}_{N}, t_{K})
	\end{array}  \right] ,
	\label{eq: scalar_snapshot_matrix}
\end{equation}
\noindent where $\boldsymbol{ \mathit{Z} } \in \mathbb{R}^{N \times K}$. Note that the scalar data are synchronized with the velocity data, that is the time instants $t_i \: (i=1 \dots K)$ in (\ref{eq: velocity_snapshot_matrix}) and (\ref{eq: scalar_snapshot_matrix}) are the same. As before, we apply singular value decomposition to the weighted matrix $\mathcal{V}^{1/2} \boldsymbol{Z}$ (where now $\mathcal{V}=diag \left(V_1, V_2 \dots V_N 
\right)$) and obtain the scalar POD modes, $U_{Z, k}(x,y,z)$, and time coefficients, $b_{k}(t)$. Thus we can write,
\begin{equation}
	c^{\prime}(x,y,z,t)
	=
	\sum^{ K }_{k=1}
	b_{k}(t)U_{Z, k}(x,y,z)
	\approx
	\sum^{m_{c}}_{k=1}
	b_{k}(t)U_{Z, k}(x,y,z),
	\label{eq:scalar_POD_expansion}
\end{equation}
\noindent where $m_{c}$ is the number of retained scalar POD modes, or in more compact form,
\begin{equation}
	\boldsymbol{c}^{\prime}(x,y,z,t)=\boldsymbol{U}_{Z}(x,y,z) \boldsymbol{b}(t).
	\label{eq:scalar_pod_coef_2}
\end{equation}
Equations \eqref{eq:velo_pod_coef_2} and \eqref{eq:scalar_pod_coef_2} can be written together as 
\begin{equation}
	\left[ \begin{array}{c}
		\boldsymbol{u}^{\prime} \\
		\boldsymbol{c}^{\prime}
	\end{array}  \right](x,y,z,t)=
	\underbrace{\left[ \begin{array}{cc}
			\boldsymbol{U}_{Y} & \boldsymbol{0} \\
			\boldsymbol{0} & \boldsymbol{U}_{Z}
		\end{array}  \right]}_{\equiv \boldsymbol{U}_{YZ}}\
	\left[ \begin{array}{c}
		\boldsymbol{a}(t) \\
		\boldsymbol{b}(t)
	\end{array}  \right].
	\label{eq:vel_scalar_pod_coef}
\end{equation}

We then build the time-delayed Hankel matrix with the POD coefficients of the most dominant $m_{u}$ velocity and $m_{c}$ scalar modes,
\begin{equation}
	\boldsymbol{H}
	=
	\left[ \begin{array}{cccc}
		\boldsymbol{a}(t_{1})& \boldsymbol{a}(t_{2})& \ldots& \boldsymbol{a}(t_{p})\\ 
		\boldsymbol{b}(t_{1})& \boldsymbol{b}(t_{2})& \ldots& \boldsymbol{b}(t_{p})\\     
		\boldsymbol{a}(t_{2})& \boldsymbol{a}(t_{3})& \ldots&
		\boldsymbol{a}(t_{p+1})\\ 
		\boldsymbol{b}(t_{2})& \boldsymbol{b}(t_{3})& \ldots&
		\boldsymbol{b}(t_{p+1})\\     
		\vdots& \vdots& \ddots& \vdots\\
		\boldsymbol{a}(t_{q})& \boldsymbol{a}(t_{q+1})& \ldots&
		\boldsymbol{a}(t_{K_{train}})\\
		\boldsymbol{b}(t_{q})& \boldsymbol{b}(t_{q+1})& \ldots&
		\boldsymbol{b}(t_{K_{train}})    
	\end{array}  \right] ,
\end{equation}

\noindent where $\boldsymbol{H} \in \mathbb{R}^{\left((m_{u}+m_{c})\times q \right) \times p}$. Performing SVD on $\boldsymbol{H}$ we obtain the matrices $\boldsymbol{\Sigma}_{H}$,  $\boldsymbol{V}_{H}$, $\boldsymbol{U}_{H}$ as before. Note that this means that we  use the same weights in the velocity and scalar coefficients; this warrants further investigation which we leave as part of future work. The matrix of the left singular vectors $\boldsymbol{U}_{H} \in \mathbb{R}^{\left((m_{u}+m_{c})\times q \right) \times r}$ can be written explicitly as, 
\begin{equation}
	\boldsymbol{U}_{H}
	=
	\left[ \begin{array}{cccc}
		\boldsymbol{U}^{(u,v,w)}_{H, 1}(t_{1})& \boldsymbol{U}^{(u,v,w)}_{H, 2}(t_{1})& \ldots&
		\boldsymbol{U}^{(u,v,w)}_{H, r}(t_{1})\\
		\boldsymbol{U}^{(c)}_{H, 1}(t_{1})& \boldsymbol{U}^{(c)}_{H, 2}(t_{1})& \ldots&
		\boldsymbol{U}^{(c)}_{H, r}(t_{1})\\    
		\boldsymbol{U}^{(u,v,w)}_{H, 1}(t_{2})& \boldsymbol{U}^{(u,v,w)}_{H, 2}(t_{2})& \ldots&
		\boldsymbol{U}^{(u,v,w)}_{H, r}(t_{2})\\ 
		\boldsymbol{U}^{(c)}_{H, 1}(t_{2})& \boldsymbol{U}^{(c)}_{H, 2}(t_{2})& \ldots&
		\boldsymbol{U}^{(c)}_{H, r}(t_{2})\\    
		\vdots& \vdots& \ddots& \vdots\\
		\boldsymbol{U}^{(u,v,w)}_{H, 1}(t_{q})& \boldsymbol{U}^{(u,v,w)}_{H, 2}(t_{q})& \ldots&
		\boldsymbol{U}^{(u,v,w)}_{H, r}(t_{q})\\
		\boldsymbol{U}^{(c)}_{H, 1}(t_{q})& \boldsymbol{U}^{(c)}_{H, 2}(t_{q})& \ldots&
		\boldsymbol{U}^{(c)}_{H, r}(t_{q})    
	\end{array}  \right] .
\end{equation}

A dynamical system for $\boldsymbol{v}_{H}$ 
\begin{equation}
	\boldsymbol{v}_{H}[k+1]
	=
	\boldsymbol{A} \boldsymbol{v}_{H}[k]
	+
	\boldsymbol{w}_{2}[k]
\end{equation}
can be derived as before. Also the process noise covariance $\boldsymbol{Q} \in \mathbb{R}^{r \times r}$ can be calculated in the same way as in section \S\ref{sec:forecast_velocity}.

Vector $\boldsymbol{v}_{H}[k]$ can be used to  obtain the current and future POD coefficients of the velocity and scalar fields as,  
\begin{eqnarray}
	\left[ 
	\begin{array}{c}
		\boldsymbol{a}[k] \\
		\boldsymbol{b}[k] \\
		\boldsymbol{a}[k+1] \\
		\boldsymbol{b}[k+1] \\
		\vdots \\
		\boldsymbol{a} [k+q] \\
		\boldsymbol{b}[k+q] \\
	\end{array} 
	\right]
	& = &  \boldsymbol{U}_{H} \Sigma_{H} 
	\boldsymbol{v}_{H}[k].
	\label{eq:PredKal_2}
\end{eqnarray}
In particular, for the $k$-th instant we have, 
\begin{equation}
	\left[ \begin{array}{c}
		\boldsymbol{a}\\
		\boldsymbol{b}
	\end{array}  \right][k]
	=
	\underbrace{\boldsymbol{U}_{H}(t_{1}) \boldsymbol{\Sigma}_{H}}
	_{=\boldsymbol{C}}
	\boldsymbol{v}_{H}[k],
\end{equation}
\noindent where $\boldsymbol{C} \in \mathbb{R}^{(m_{u}+m_{c}) \times r}$ and matrix $\boldsymbol{U}_{H}(t_{1})$ represents the top two rows of $\boldsymbol{U}_{H}$, i.e.\ 
\begin{equation}
	\boldsymbol{U}_{H}(t_{1})
	=
	\left[ \begin{array}{cccc}
		\boldsymbol{U}^{(u,v,w)}_{H, 1}(t_{1})& \boldsymbol{U}^{(u,v,w)}_{H, 2}(t_{1})& \ldots&
		\boldsymbol{U}^{(u,v,w)}_{H, r}(t_{1})\\
		\boldsymbol{U}^{(c)}_{H, 1}(t_{1})& 
		\boldsymbol{U}^{(c)}_{H, 2}(t_{1})& \ldots&
		\boldsymbol{U}^{(c)}_{H, r}(t_{1})
	\end{array}  \right] .
	\label{eq:U_H(u,c)_detail}
\end{equation}

Let's assume that we have now $l$ scalar measurements  $\boldsymbol{s}[k]$; they can written as 
\begin{equation}
	\boldsymbol{s}[k]
	=
	\boldsymbol{S} \boldsymbol{U}_{Z} \boldsymbol{b}[k]
	+
	\boldsymbol{g}[k],
\end{equation}
\noindent where now matrix $\boldsymbol{S}$ selects the rows of the POD mode matrix $\boldsymbol{U}_{Z}$ corresponding to the scalar sensor locations. Note that it is possible to mix velocity and scalar measurements; in this case $\boldsymbol{S}$ will act on the compound matrix $\boldsymbol{U}_{YZ}$, see \eqref{eq:vel_scalar_pod_coef}. In the following we assume that we have scalar measurements only. The covariance $\boldsymbol{R}$ of  vector $\boldsymbol{g}$ can be obtained as explained in the previous section. 

The Kalman filter takes the form,
\begin{subeqnarray}
	\hat{\boldsymbol{v}}_{H}[k+1]
	& = &
	\boldsymbol{A} \hat{\boldsymbol{v}}_{H}[k]
	+
	\mathcal{L} \big(
	\boldsymbol{s}[k]-\hat{\boldsymbol{s}}[k]
	\big),\\
	\left[ \begin{array}{c}
		\hat{\boldsymbol{a}}\\
		\hat{\boldsymbol{b}}
	\end{array}  \right] [k]
	& = &
	\boldsymbol{C} \hat{\boldsymbol{v}}_{H}[k],\\
	\hat{\boldsymbol{s}}[k]
	& = &
	\boldsymbol{S}  \boldsymbol{U}_{Z}  \hat{\boldsymbol{b}}[k]
	=
	\boldsymbol{S}  \boldsymbol{U}_{Z}  (\boldsymbol{C})_2
	\hat{\boldsymbol{v}}_{H}[k].     
	\label{eq:random_KL}
\end{subeqnarray}
\noindent where $(\boldsymbol{C})_2$ indicates the second row block of matrix $\boldsymbol{C}$, and the Kalman gain matrix $\mathcal{L}$ is obtained by solving a Riccati equation similar to \eqref{eq:Riccati} where $\boldsymbol{U}_{Y}$ is replaced by $\boldsymbol{U}_{Z}$. 

From $\hat{\boldsymbol{v}}_{H}[k]$ we can estimate the current and future POD coefficients from \eqref{eq:PredKal_2} and the instantaneous velocity and scalar fields from \eqref{eq:vel_scalar_pod_coef}.

\section{Application to the flow around a surface-mounted cube}\label{sec:flow_surf_mounted_cube}

\subsection{Computational set-up and numerical methodology}
\label{sec:JFM_CompSetUp}
\begin{figure}[h!]
	\centerline{\includegraphics[width=\linewidth]{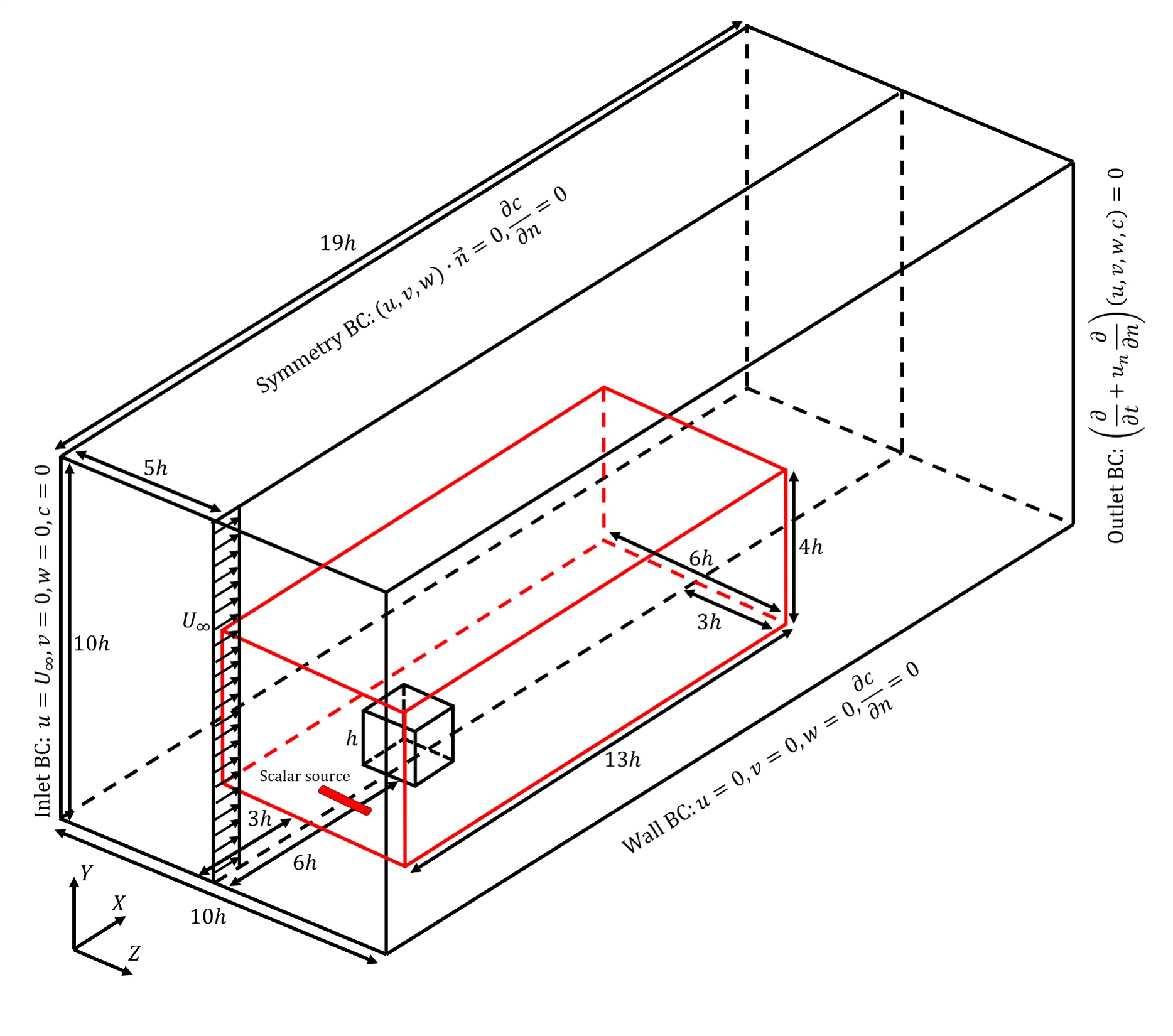}}
	\caption{Computational domain and boundary conditions. The red lines mark the boundaries of the region where snapshot data are collected.}
	\label{fig:01_Domain}
\end{figure}

We consider the three-dimensional flow around a surface-mounted cube of height $h$. The computational domain, shown in figure \ref{fig:01_Domain}, has dimensions $L_{x} \times L_{y} \times L_{z}= 19h \times 10h \times 10h$. The origin of the coordinate system is located at the bottom mid point of the upstream face of the cube. The inlet is located at $x/h=-6$ in the streamwise direction and the outlet at $x/h=13$.  The domain extends between $-5 \leq z/h \leq 5$ in the spanwise direction and up to $y/h=10$ in the wall-normal direction. Uniform velocity $U_{\infty}=1$ is prescribed at the inlet and a convective boundary condition at the outlet. No-slip conditions are imposed on the cube surfaces and bottom wall, while symmetry conditions are applied on the top and spanwise boundaries \citep{Krajnovic2002}. 

Scalar is released from a source with elliptical cross-section centered at $(x_{s}, y_{s}, z_{s})=(-2, 0.1, -0.5:+0.5)h$. The axis of the source is along the $z$ direction and at the same elevation as the core of the horseshoe vortex forming in front of the cube (see later). The source strength $\hat{m}_{T}(\boldsymbol{x})$ (amount of scalar released per unit volume) is steady with a spatial distribution given by
\begin{equation}
	\hat{m}_{T}=
	2\gamma 
	\left[ cos(r\pi)+1 \right],
	\label{eq:Sca_CosLaw}
\end{equation}
with 
\begin{equation}
	\gamma
	=  \frac{1}{4r_{x}r_{y}(\pi-2/{\pi}) }, \quad 
	r
	=  \sqrt{
		\bigg( \frac{x-x_{s}}{r_{x}} \bigg)^{2}
		+
		\bigg( \frac{y-y_{s}}{r_{y}} \bigg)^{2} 
	},
	\label{eq:Sca_CosLaw_gamma}
\end{equation}
\noindent where $(x_{s}, y_{s}, z_{s})$ are the coordinates of the center of the source, $r_{x}=0.1h$ is the major axis, and 
$r_{y}=0.08h$ is the minor axis (smaller than $r_x$ due to the presence of the bottom wall). Equation \eqref{eq:Sca_CosLaw} represents a source distribution that vanishes at the boundary of the elliptical cross-section. The normalization parameter $\gamma$ ensures the volume integral of $\hat{m}_{T}$ is unity. 

The flow is simulated with the in-house code PANTARHEI. The incompressible Navier-Stokes equations are discretised is space using the finite volume method, with a second order central discretization scheme (for both convection and viscous terms) and marched in time with a 3rd order backward scheme. The fractional step method is employed to obtain pressure and correct the velocities to satisfy the continuity equation. The code has been used extensively to simulate transitional and turbulent flows   \citep{Yao_papadakis_2023,Schlander_Rigopoulos_Papadakis_2024,Thomareis_Papadakis_2017,Thomareis_Papadakis_2018}. 

The Reynolds number, defined as $Re_{h}=U_{\infty}h/\nu$, is set to $5000$. The flow domain is discretized using a Cartesian mesh. The cells are clustered close to the cube surfaces and bottom wall. The time step $\delta t$ is selected to satisfy $CFL<0.5$.  More details are provided in table \ref{tab:JFM_MeshSpec}. The maximum ratio of the cell size (taken as the  cubic root of cell volume) to the Kolmogorov length scale was equal to 4.6 in the recirculation zone behind the cube, which indicates almost DNS-quality resolution sufficient for the purposes of the present investigation. 

Snapshots of velocities and scalar fields are collected within a sub-region (defined by the red lines of figure \ref{fig:01_Domain}) that contains $N=38.69 \times 10^{6}$ cells. The flow is first advanced for $2$ flow-through times, the simulation is then restarted and advanced for $6$ more flow-through times. In total $K=6000$ flow and scalar field snapshots are recorded synchronously during the last $6$ flow-through times. The time separation between snapshots is $\Delta t=0.019{h}/{U_{\infty}}$. 

\begin{table}[h!]
	\begin{center}
		\def~{\hphantom{0}}
		\begin{tabular}{lcccc}
		Total No. of cells & $N_{x} \times N_{y} \times N_{z}$ & $N_{\text{cube}}$ & $\delta_{\text{1st}}/h$ & $U_{\infty} \delta t/h$ \\[3pt] 
		61 612 200 & $576 \times 268 \times 406$ & $102$ & $0.006$ & $0.001$ 	
		\end{tabular}
		\caption{Computational details. $N_{x}, N_{y}, N_{z}$ are the number of cells in the streamwise, wall-normal and spanwise directions respectively, $N_{\text{cube}}$ is the number of subdivisions along the cube edge, $\delta_{\text{1st}}$ is the thickness of the first layer of cells near the walls, and $\delta t$ the time step.}
		\label{tab:JFM_MeshSpec}
	\end{center}
\end{table}

\subsection{Time-averaged flow and scalar fields}

\begin{figure}[h!]
	\centerline{\includegraphics[width=0.94\linewidth]{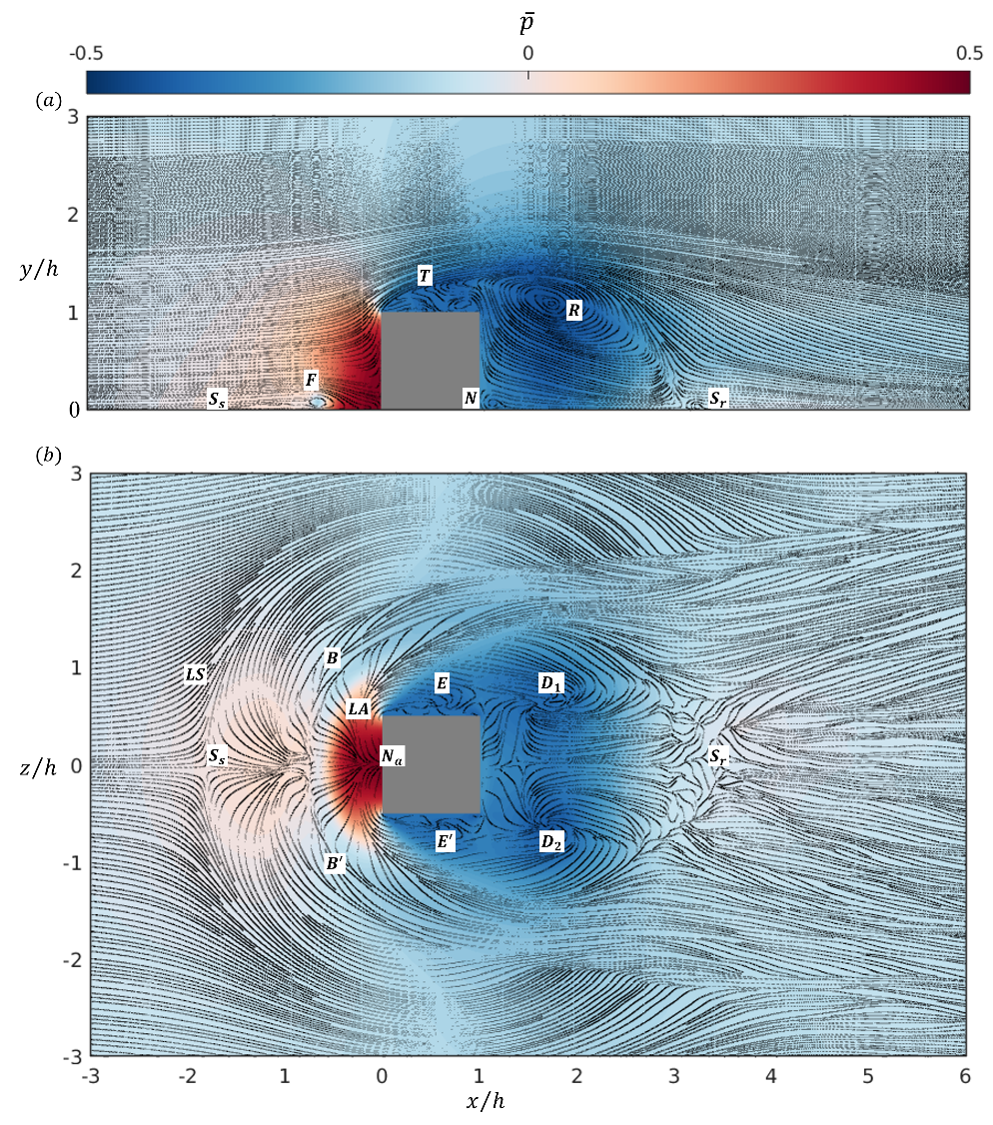}}
	\caption{Time-averaged streamlines superimposed on contours of mean pressure field: ($a$) symmetry $xy$-plane at $z/h=0$, (b) $xz$-plane at distance $y/h=0.003$ from the bottom wall.}
	\label{fig:07_XYXZ_PStreamlines}
\end{figure}

Figure \ref{fig:07_XYXZ_PStreamlines} shows time-averaged streamlines superimposed on the mean pressure fields on the symmetry $xy$-plane and the $xz$-plane at the height of the first cell centroid away from the bottom wall. The reference pressure is the mean static pressure at the inlet plane. In panel ($a$), three main separation regions can be seen, at the front ($F$), at the  leading edge ($T$) and downstream of the cube ($R$). There is a secondary recirculation region ($N$) at the bottom corner of the leeward face. 

In panel ($b$), it can be seen that the flow separates upstream of the cube, at the saddle point $S_s$ located at $(-1.8, 0, 0)h$. The flow reattaches upstream of the cube at the nodal point $N_{a}$ at $(-0.015, 0, 0)h$. The streamline passing through $S_{s}$ bends around the cube and forms a "line of separation" ($LS$) close to the bottom wall. The "line of attachment" $LA$ is formed by streamlines passing through $N_{a}$. The horseshoe vortex center forms a line between $LS$ and $LA$ and is deflected around the cube to form two legs $B$ and $B^{\prime}$. The shear layers separating from the two vertical edges of the front face generate the two lateral vortices $E$ and $E^{\prime}$. The two vortices and the shear layer over the cube join at a higher elevation to form an arc-shaped vortex tube. Downstream of the cube, two symmetrically located points $D_{1}$ and $D_{2}$ indicate two vortices on the bottom wall. Further downstream, the flow reattaches at $S_r$ $(3.31, 0, 0)h$. The reattachment length is $L_{R}/h=2.31$. 

\begin{figure}[h!]
	\centerline{
		\includegraphics[width=0.7\columnwidth]{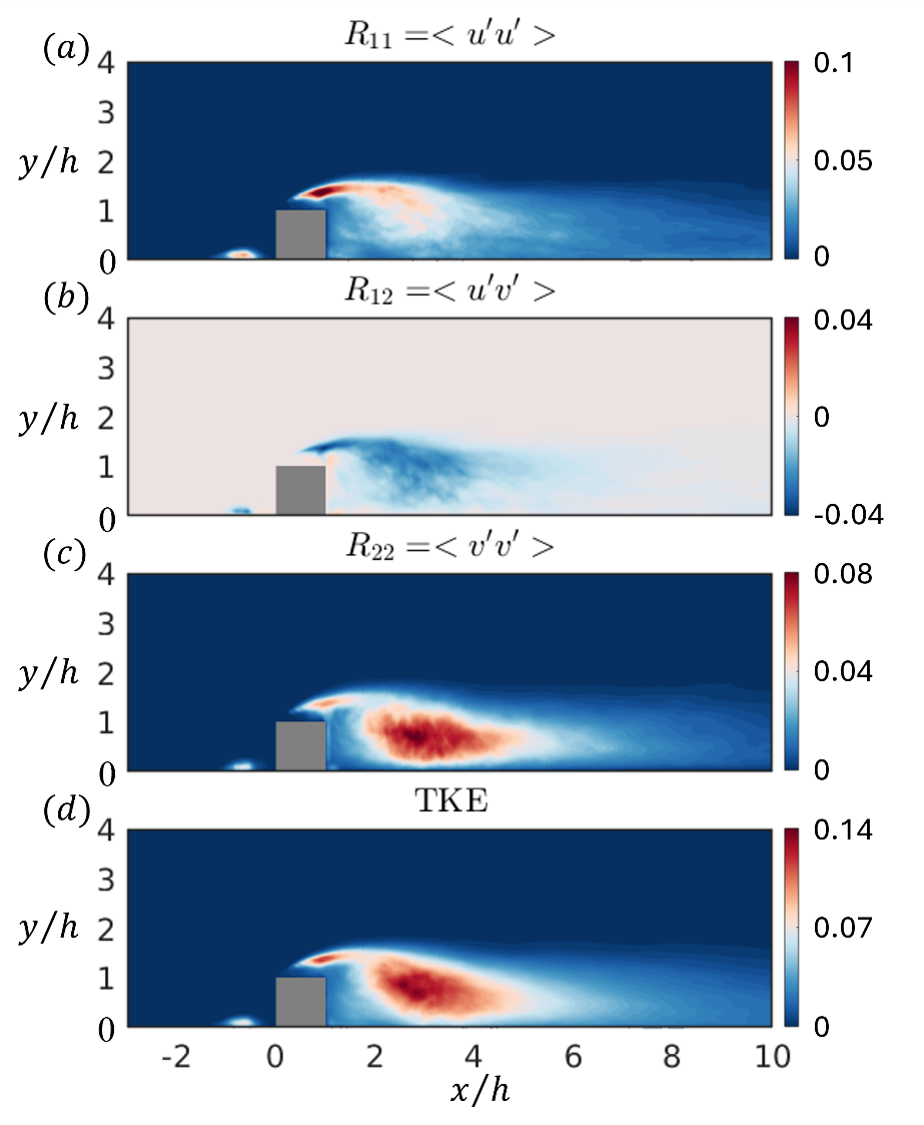}}
	\caption{Contours of the Reynolds stresses and turbulent kinetic energy (TKE) at symmetry $xy$-plane at $z/h=0$.}
	\label{fig:09_XY0_Rexx}
\end{figure}

Contour plots of the Reynolds stresses and the turbulent kinetic energy (TKE) at the symmetry plane  $z/h=0$ are shown in figure \ref{fig:09_XY0_Rexx}. High levels of $\langle u^{\prime}u^{\prime} \rangle$ are found inside the separating shear layer emanating from the leading edge. The peak value is located above the cube at $(x,y)=(0.9, 1.36)h$, while $\langle v^{\prime}v^{\prime} \rangle$ peaks downstream of the cube at $(x,y)=(2.85, 0.71)h$. The peak of $\langle u^{\prime}v^{\prime} \rangle$ is located at $(x,y)=(0.95, 1.36)h$ which is close to the peak of $\langle u^{\prime}u^{\prime} \rangle$. Finally, the TKE peak is located at $(x,y)=(2.6, 0.86)h$ closer to the peak location of $\langle v^{\prime}v^{\prime} \rangle$. Small patches of Reynolds stresses can be seen upstream of the cube, around the horseshoe vortex center.

\begin{figure}[h!]
	\centerline{
		\includegraphics[width=0.7\columnwidth]{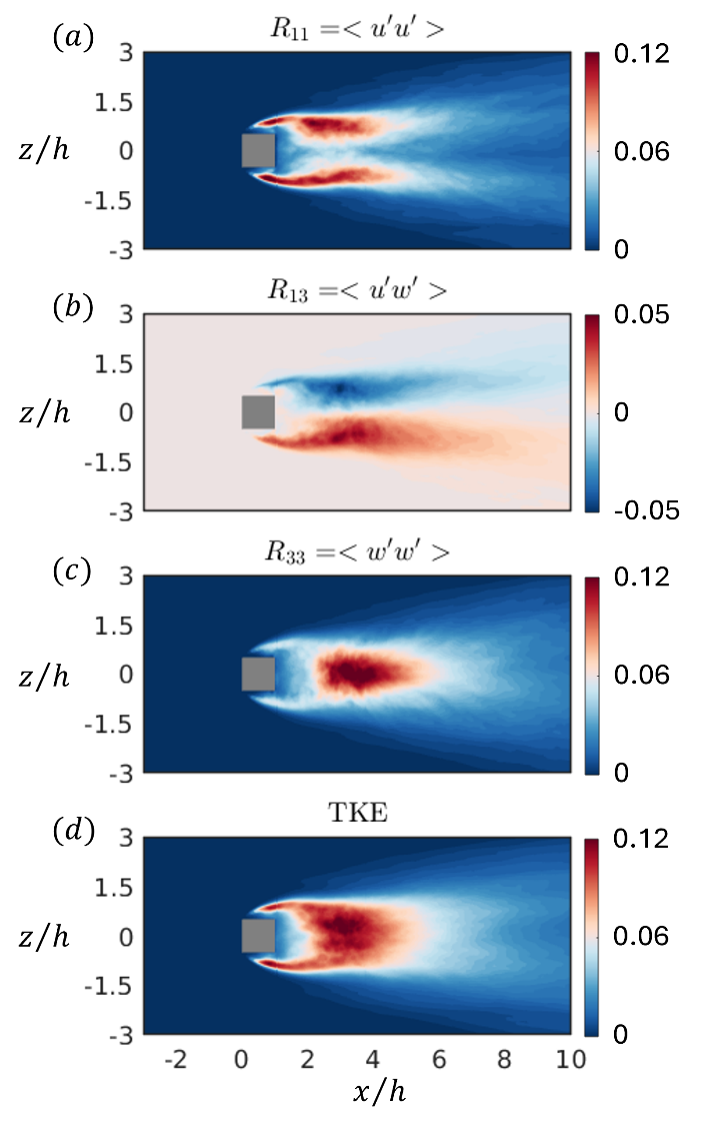}}
	\caption{Contours of the Reynolds stresses and TKE in $xz$-plane at mid-height $y/h=0.5$.}
	\label{fig:10_XZ0D5_Rexx}
\end{figure}

Figure \ref{fig:10_XZ0D5_Rexx} shows contours of Reynolds stresses and 
and TKE in the $xz$-plane at mid-height $y/h=0.5$. High levels of $\langle u^{\prime}u^{\prime} \rangle$ and $\langle u^{\prime}w^{\prime} \rangle$ are found inside the shear layers separating from the front vertical edges and downstream of the cube. The peak value of $\langle u^{\prime}u^{\prime} \rangle$ is found at $(x,z)=(0.8, \pm 0.85)h$. Notice the high values of the normal stresses in the spanwise direction $\langle w^{\prime}w^{\prime} \rangle$ that peak further downstream at $(x,z)=(3.15, 0)h$. Such high values indicate strong symmetry-breaking motions. The TKE combines features of $\langle u^{\prime}u^{\prime} \rangle$ and $\langle w^{\prime}w^{\prime} \rangle$ but is more influenced by the spanwise stresses.

\begin{figure}[h!]
	\centerline{
		\includegraphics[width=\linewidth]{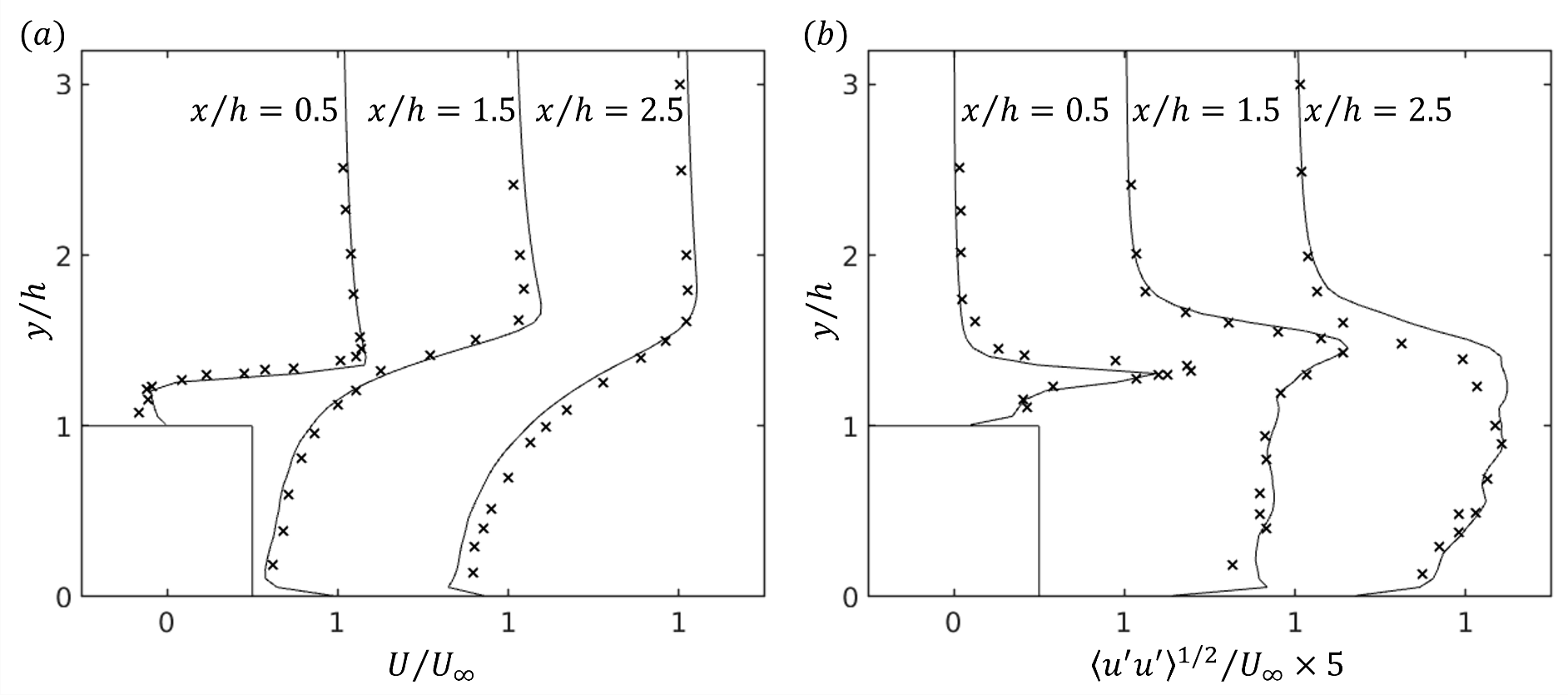}}
	\caption{Comparison of numerical results with measurements of mean and rms of streamwise velocity in symmetry $xy$ plane at $z/h=0$: $-$ present DNS, $\times$ \cite{Castro1977UniFlow}.}
	\label{fig:12_XY0_ExpUm}
\end{figure}

Mean and rms profiles of the streamwise velocity are compared against the experimental data of \cite{Castro1977UniFlow} in Fig. \ref{fig:12_XY0_ExpUm}. At $x/h=0.5$ and $1.5$ the DNS solution follows the experimental results very well. At $x/h=2.5$, the numerical results are also good, but slightly overestimate the backflow streamwise velocity  within the recirculation zone in $y/h=0-1$. For the rms of streamwise velocity, the numerical results give quantitatively good predictions at all streamwise positions. 

\begin{figure}[h!]
	\centerline{
		\includegraphics[width=\linewidth]{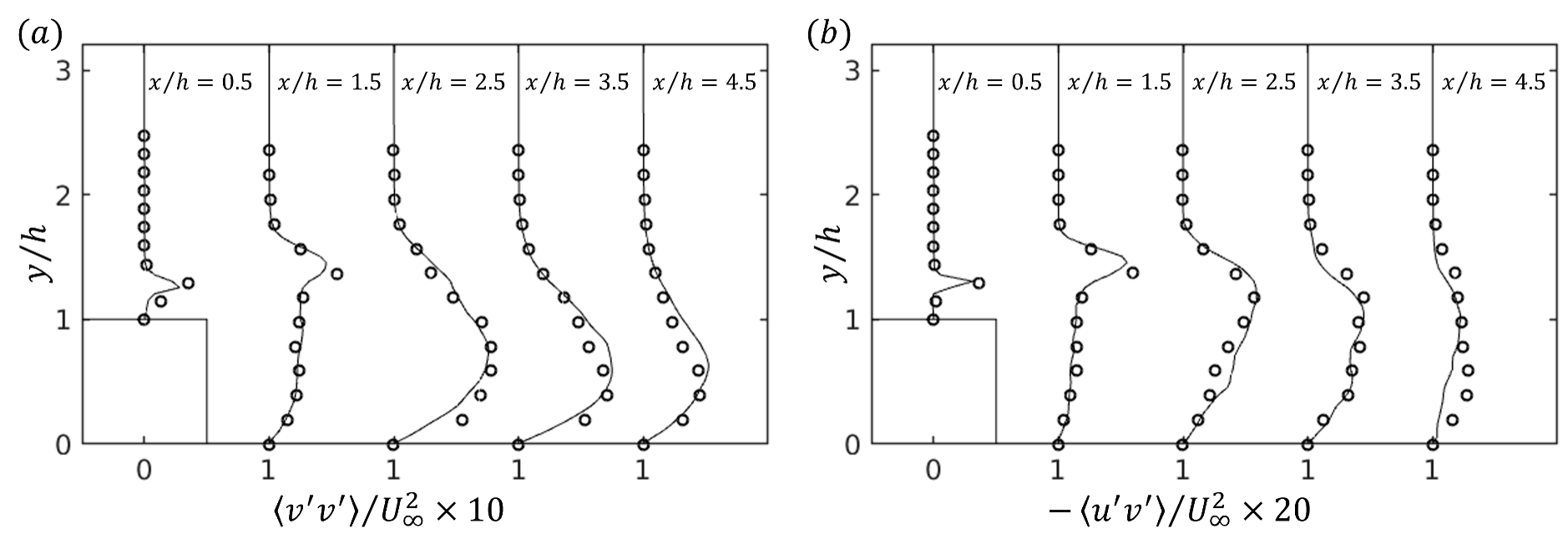}}
	\caption{Comparison of present results (solid lines) for Reynolds stresses at the symmetry $xy$ plane at $z/h=0$ against the DNS results (circles) of \cite{Rossi-2010}.}
	\label{fig:13_XY0_ExpRexx}
\end{figure}
We also compare our predictions to the DNS of \cite{Rossi-2010}. Figure \ref{fig:13_XY0_ExpRexx} presents profiles of normal stresses $\langle v^{\prime}v^{\prime} \rangle$, $\langle u^{\prime}v^{\prime} \rangle$. The present results compare very well with the literature at all locations. 

\begin{figure}[h!]
	\centerline{
		\includegraphics[width=0.8\linewidth]{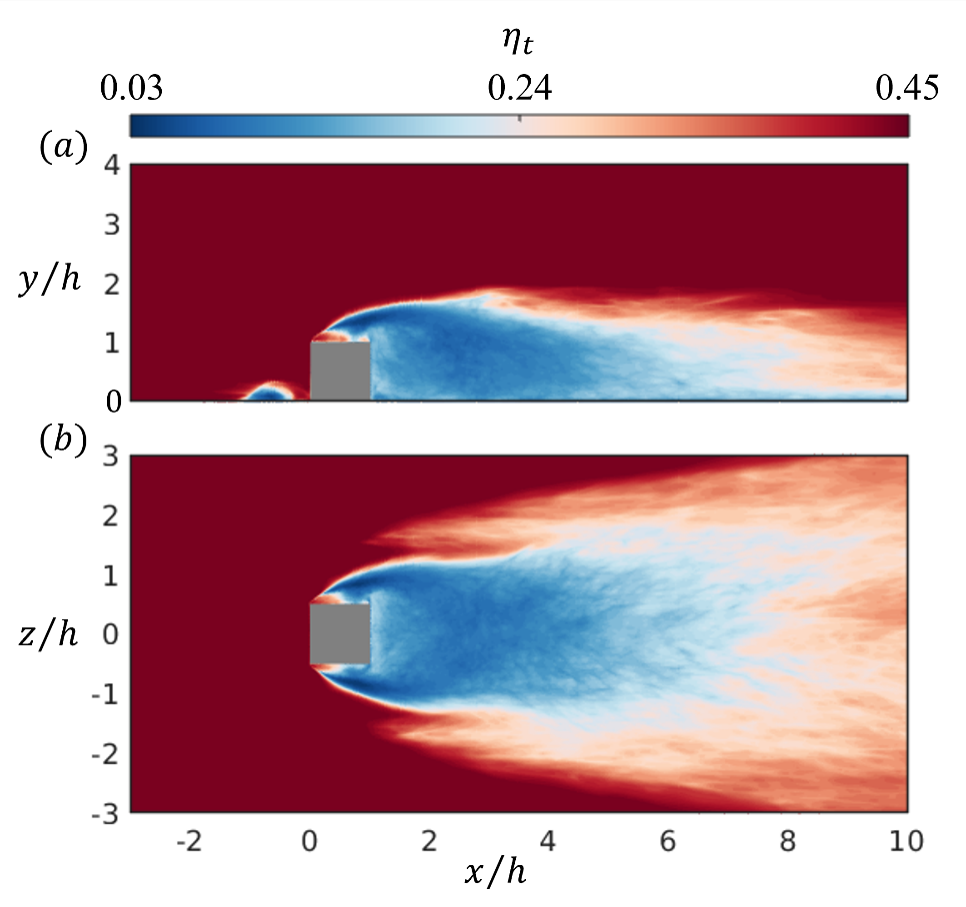}
	}
	\caption{Contour plots of the Kolmogorov time scale $\eta_{t}$ ($a$) symmetry $xy$-plane at $z/h=0$, ($b$) $xz$-plane at mid-height $y/h=0.5$.}
	\label{fig:03_XYXZ_Kolomt}
\end{figure}

Contours of the Kolmogorov time scale $\eta_{t}= \sqrt{\frac{\nu}{\langle \epsilon \rangle +0.001}}$, where $\nu$ is the kinematic viscosity and $\langle \epsilon \rangle$ the mean dissipation rate, are shown in figure \ref{fig:03_XYXZ_Kolomt} at two planes. The areas with meaningful values are in the separating shear layers and inside the recirculation zone. The $\eta_{t}$ values range between $0.03-0.2$ time units (one time unit is equal to $h/U_\infty$). It is expected that the time scale associated with the maximum Lyapunov exponent will of similar magnitude, as explained in the Introduction.

\begin{figure}[h!]
	\centerline{
		\includegraphics[width=0.6\linewidth]{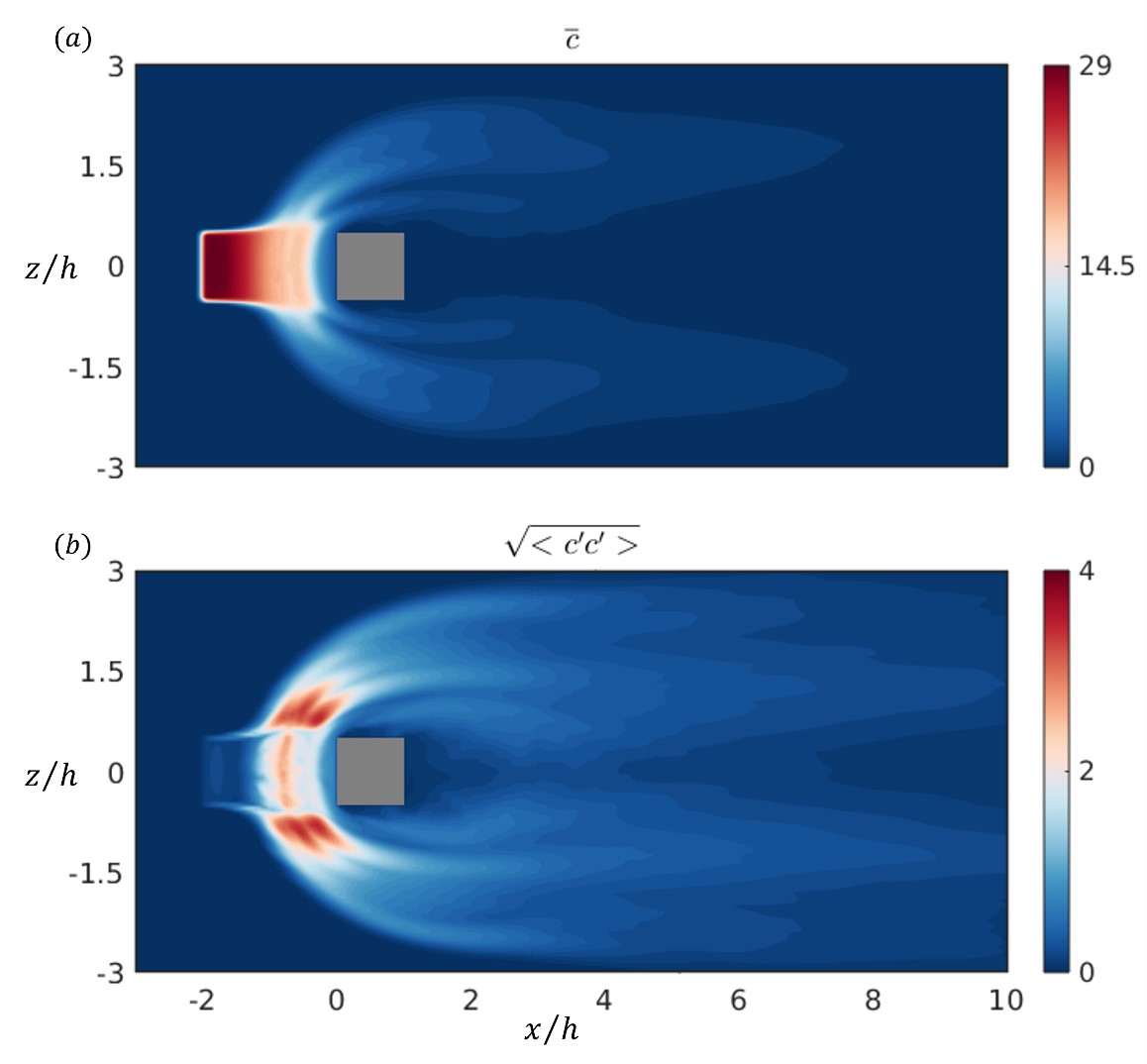}
	}
	\caption{Contour plots of the ($a$) mean scalar field and ($b$) rms of scalar field in $xz$ plane at $y/h=0.1$ (height of the source center).}
	\label{fig:15_XZ0D1_Cm}
\end{figure}

Contours of the mean and rms of scalar on the $xz$-plane at $y/h=0.1$ are shown in figure \ref{fig:15_XZ0D1_Cm}. The scalar is convected towards the cube, bends around it,  and due to dilution effect the concentration drops. The scalar dispersion is dominated by the horseshoe vortex near the cube. The scalar rms values are maximised in the region where the flow around the cube bends. This is the area of the highest mean scalar gradient (see top panel) and turbulent generation due to the separating shear layers from the vertical edges of the front face.

\subsection{POD modes}
\begin{figure}[h!]
	\centerline{
		\includegraphics[width=\linewidth]{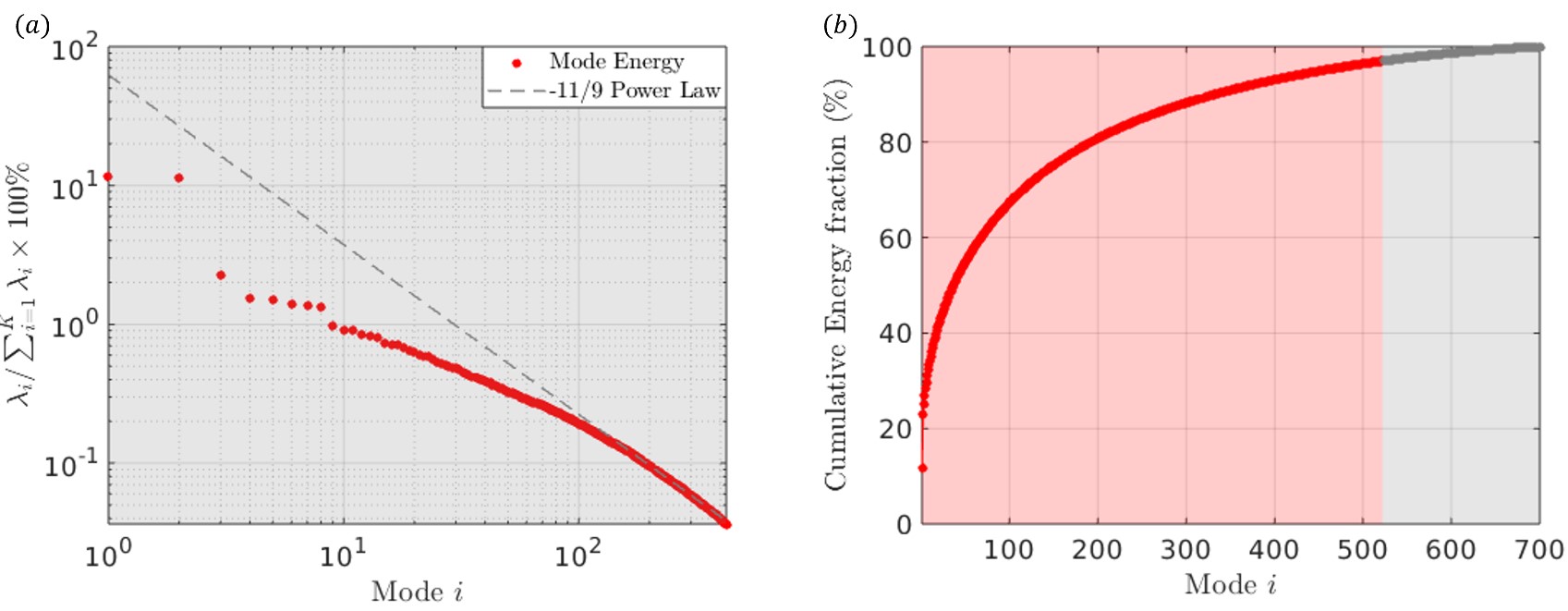}}
	\caption{Energy fraction of ($a$) the first $424$ velocity POD modes and ($b$) the cumulative energy.}
	\label{fig:19_UVWLambda}
\end{figure}

We implemented the sequential SVD method of \citet{Li2021} to obtain the POD eigenvectors and time coefficients from  $K=6000$ velocity and scalar snapshots. This method reduces significantly the memory requirements. The left panel of figure \ref{fig:19_UVWLambda} shows the energy fraction of each mode in log scale. It can be seen that the first two modes have the same energy (about $10\%$ of the total) and they are paired (as will be seen later). Modes $3$-$8$ have similar energy (between $1-3\%$), all other modes have energy less than $1\%$. 

It is very interesting to note that the energy distribution of the higher modes (with index larger than 100) follows a power law with slope $-11/9$. This power law  was first derived theoretically in \cite{knight1990} from the well-known $\kappa^{-5/3}$ law for the energy density with respect to wavenumber $\kappa$ in the inertial regime for homogeneous isotropic turbulence. The fact that the energy follows the $-11/9$ power law confirms that the turbulent flow is well resolved. The right plot displays the cumulative energy, the first $523$ modes (shaded area) account for $97\%$ of the total energy. 

\begin{figure}[h!]
	\centerline{
		\includegraphics[width=\linewidth]{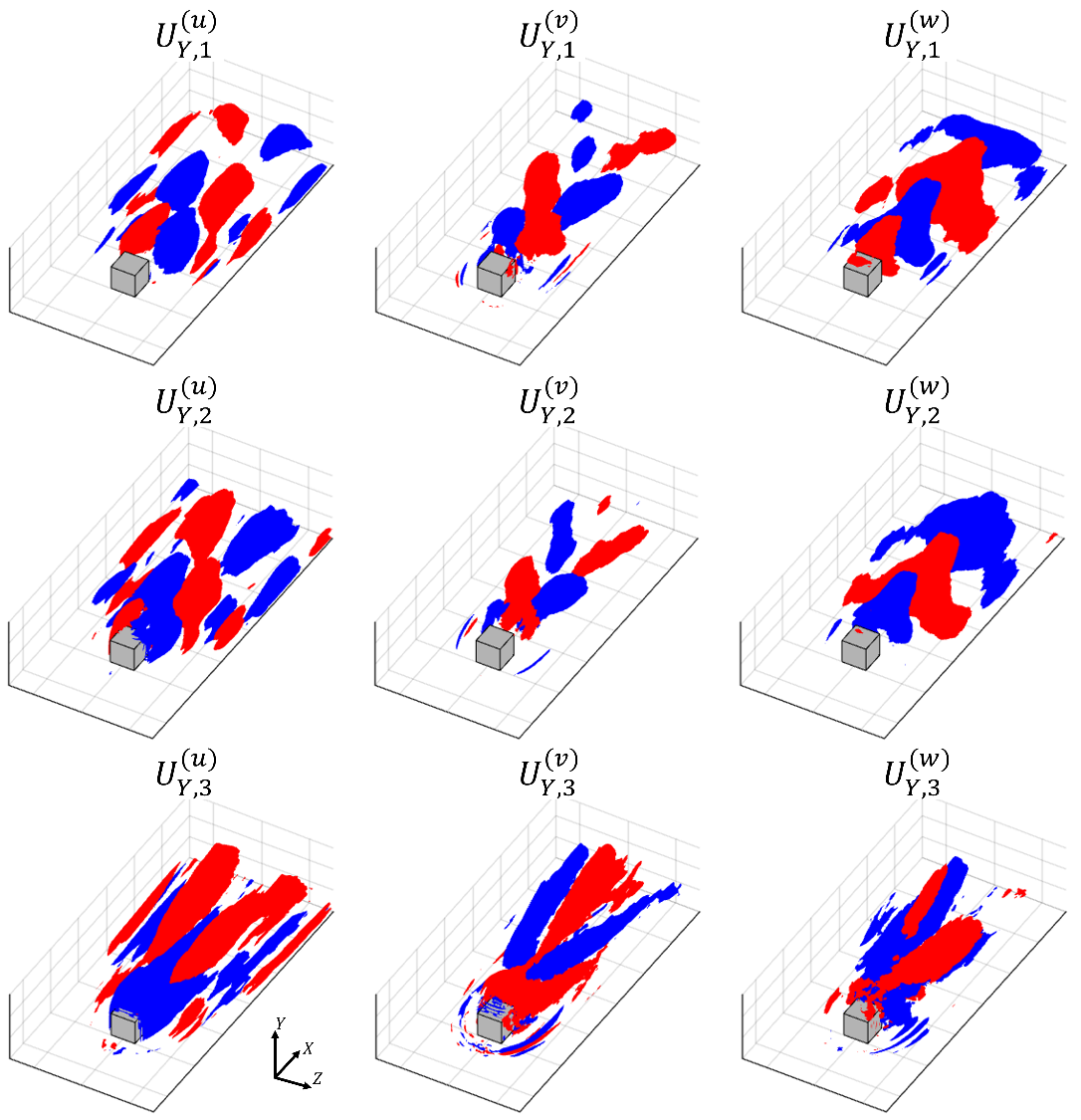}}
	\caption{Iso-surfaces of (left column) $\phi_{u, i}$, (middle column) $\phi_{v, i}$, (right column) $\phi_{w, i}$ for POD modes $1-3$. Iso-surfaces of the velocity modes are normalized with the $L_{\infty}$-norm: $U_{:,1}$ and $U_{:,2}$ blue (-0.2), red (+0.2); $U_{:,3}$ blue (-0.16), red (+0.16).}
	\label{fig:20_UVWPhi01-03}
\end{figure}

Iso-surfaces of the three dominant modes are shown in Fig. \ref{fig:20_UVWPhi01-03}. The streamwise modes $U^{(u)}_{Y,1}$ and $U^{(u)}_{Y,2}$ illustrate three-dimensional structures that are antisymmetric with respect to the $xy$ $z/h=0$ plane and are shed downstream in alternating fashion. Similar anti-symmetric and alternating behavior can also be observed for the wall-normal modes $U^{(v)}_{Y,1}$ and $U^{(v)}_{Y,2}$. The spanwise velocity modes ($U^{(w)}_{Y,1}$ and $U^{(w)}_{Y,2}$) on the other hand consist of distinct structures which are symmetric and are spatially shifted in the streamwise direction. This spatial shift combined with the temporal shift between the POD coefficients (not shown) and the sharp spectral peak at the same frequency as will be shown later results in downstream propagating structures. The eigenvectors of the streamwise and wall-normal velocities of the third mode are symmetric and consist of shear-layer structures originating from the upstream edges of the cube. Moreover, a single pair of counter-rotating streamwise structures is also observed in iso-surfaces of $U^{(u)}_{Y,3}$ and $U^{(v)}_{Y,3}$ downstream of the cube, see also \cite{Bourgeois_et_al_2011}. Higher order modes consist of smaller structures that are difficult to visualise and interpret. 

\begin{figure}[h!]
	\centerline{\includegraphics[width=\linewidth]{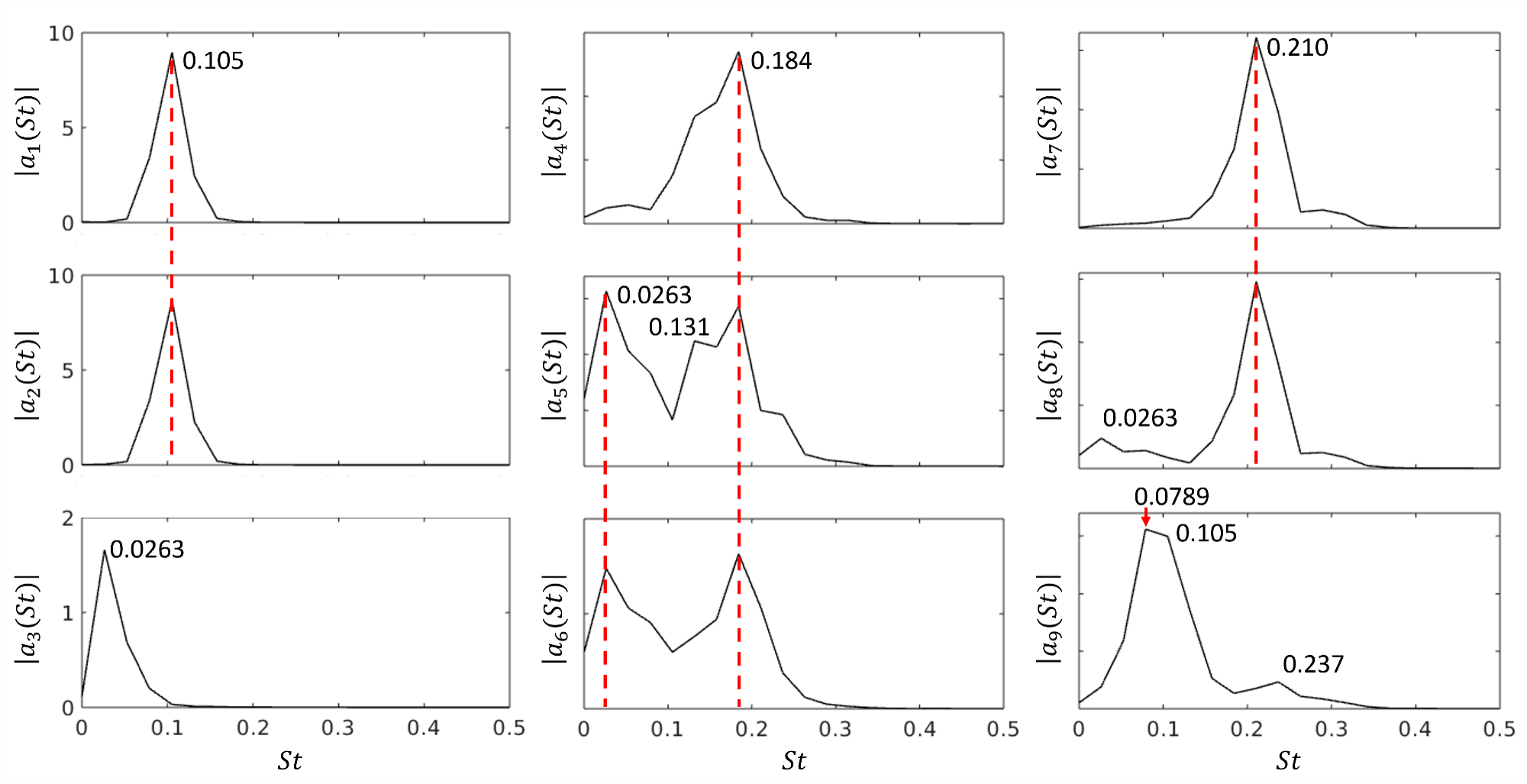}}
	\caption{Spectra of time coefficients $a_{i}(t)$ of the $9$ most dominant POD modes.}
	\label{fig:23_UVWSt}
\end{figure}

The spectra of the POD coefficients of the 9 most dominant modes are shown in figure \ref{fig:23_UVWSt}. The spectra were obtained using the Hanning windowing method. The time signal was split in $5$ segments with an overlapping ratio of $50\%$. The first two modes peak at the same frequency, $St \approx 0.105$. A low-frequency peak $St=0.0263$ with high intensity is found in the spectrum of the third mode. This low-frequency peak persists in the spectra of modes $5$ and $6$. Mode $4$ peaks at $St=0.184$ ($\approx 2\times 0.105-0.0263$, i.e.\ a linear combination of the previous two frequencies). The first harmonic of modes $1$ and $2$ appears in the spectra of modes $7$ and $8$. 

\begin{figure}[h!]
	\centerline{
		\includegraphics[width=\linewidth]{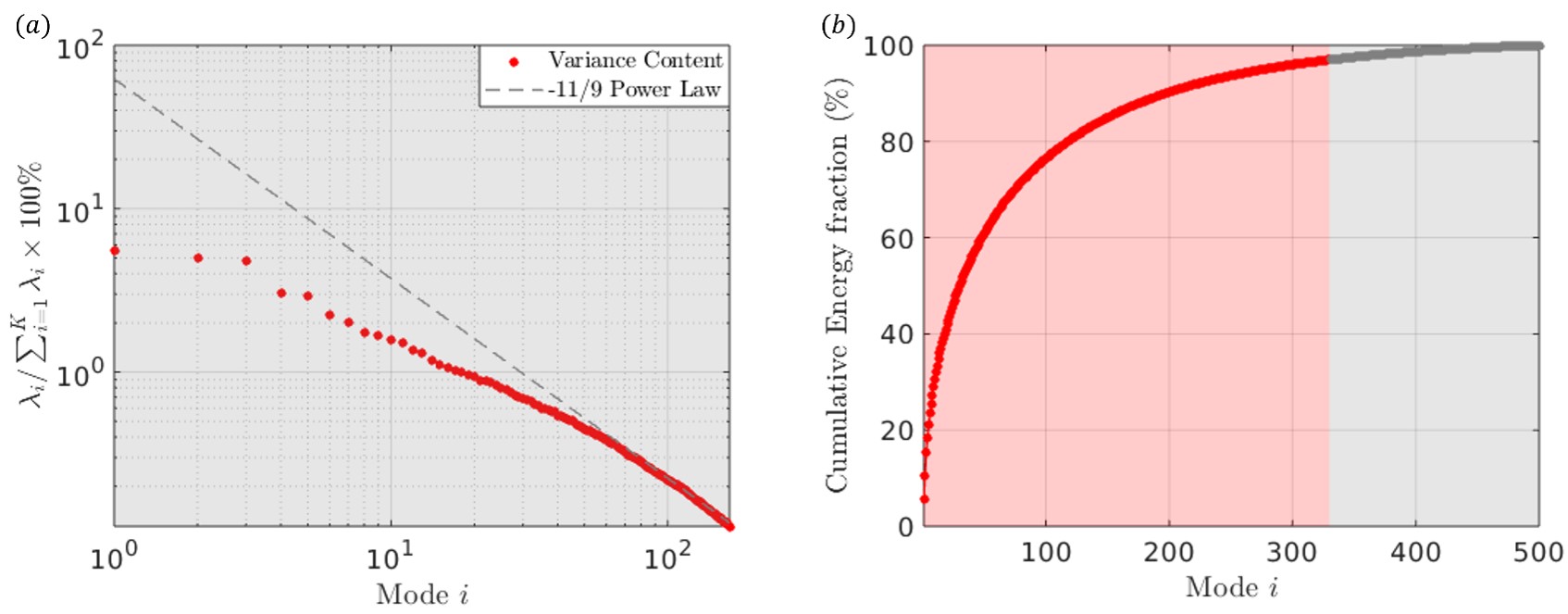}}
	\caption{Variance fraction of ($a$) the first $167$ scalar POD modes and ($b$) the cumulative variance: snapshots number $K=6000$.}
	\label{fig:24_SCALambda}
\end{figure}

Figure \ref{fig:24_SCALambda} shows the variance fraction of the scalar POD modes. It also follows the $-11/9$ power law for mode numbers larger than $\approx 70$. The first $330$ modes (shown in red-shaded area) account for $97\%$ of scalar variance. Thus fewer modes are needed for the scalar compared to the flow field to capture the same percentage of the variance. 

\begin{figure}[h!]
	\centerline{
		\includegraphics[width=\linewidth]{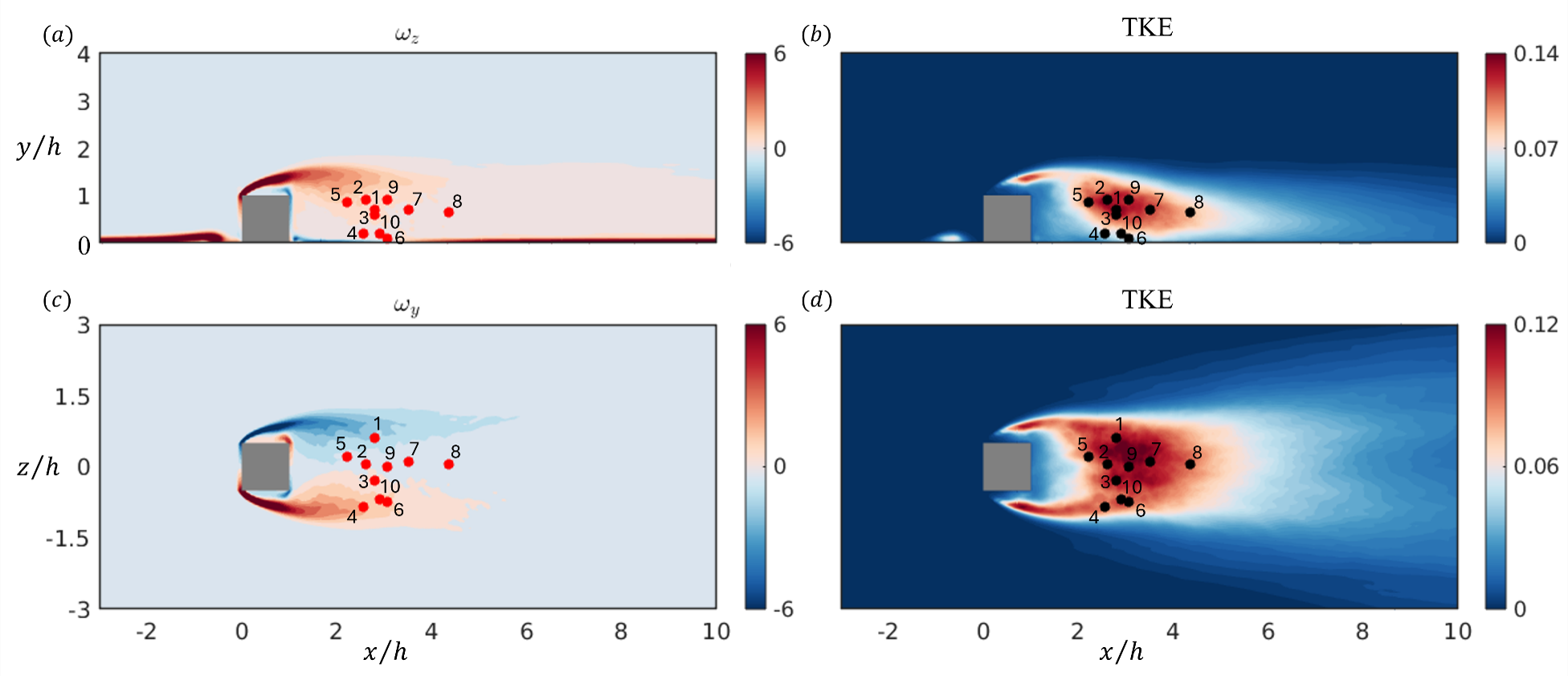}}
	\caption{Locations of the first $10$ sensors placed at the peaks of the $10$ most dominant velocity POD modes superimposed on contours of spanwise ($a$) and wall-normal vorticity ($b$) and TKE at the symmetry $xy$ plane ($a$ \& $b$) and $xz$ plane at $y/h=0.5$ ($c$ \& $d$).}
	\label{fig:18_VORTKEsensors}
\end{figure}

Sensors are placed at the peaks of the velocity POD modes. The sensors measure the three velocity fluctuations $u^{\prime}(t)$, $v^{\prime}(t)$ and $w^{\prime}(t)$. The sensor locations from the leading $10$ modes superimposed on contours of the mean vorticity and turbulent kinetic energy are shown in figure \ref{fig:18_VORTKEsensors}. As intuitively expected, the sensors are clustered in the region of high turbulent kinetic energy. They are also symmetrically distributed about the $xy$-plane at $z/h=0$ in the wake. Scalar sensors are also placed at same locations. 

\subsection{Flow field reconstruction and forecasting from velocity measurements} \label{subsec:VelOnly}
We first use $q=1$ and assemble the Hankel matrix using time coefficients of the first $m_{u}$ velocity POD modes. The parameter $m_{u}$ is varied from $18, 37$ to $68$ accounting for $40\%, 50\%$ and $60\%$ of the turbulent kinetic energy content respectively. We use $K_{train}=4000$ snapshots to extract the model matrices (training dataset) and the rest of the snapshots (i.e.\ $2000$) to validate the estimator (validation dataset). The reconstruction quality is quantified with the FIT[$\%$] metric which is defined as, 
\begin{equation}
	\text{FIT}[\%]
	=100 
	\left (1-
	\frac{ 
		\sum^{10}_{i=1} 
		\overline{ \left( {a_{i}[k]}-\hat{a}_{i}[k] \right)^{2} }  
	}
	{ 
		\sum^{10}_{i=1} 
		\overline{a^{2}_{i}[k]} 
	}
	\right), 
	\label{eq:rms_of_ae}
\end{equation}
\noindent where the overbar denotes average over the index $k$ in the validation dateset. This metric quantifies the difference between the true and estimated time coefficients for the first $10$ POD modes (that contain $35\%$ of the total energy). The flow is turbulent and hundreds of POD modes are needed to capture the kinetic energy of the system, as shown in  figure \ref{fig:19_UVWLambda}(b). Higher order modes have narrow spatial footprint, it is thus unrealistic to expect that a small number of sensors will be able to reconstruct the total energy. For this reason, we target the energy of the $10$ most dominant modes.

\begin{figure}[h!]
	\centerline{
		\includegraphics[width=0.4\linewidth]{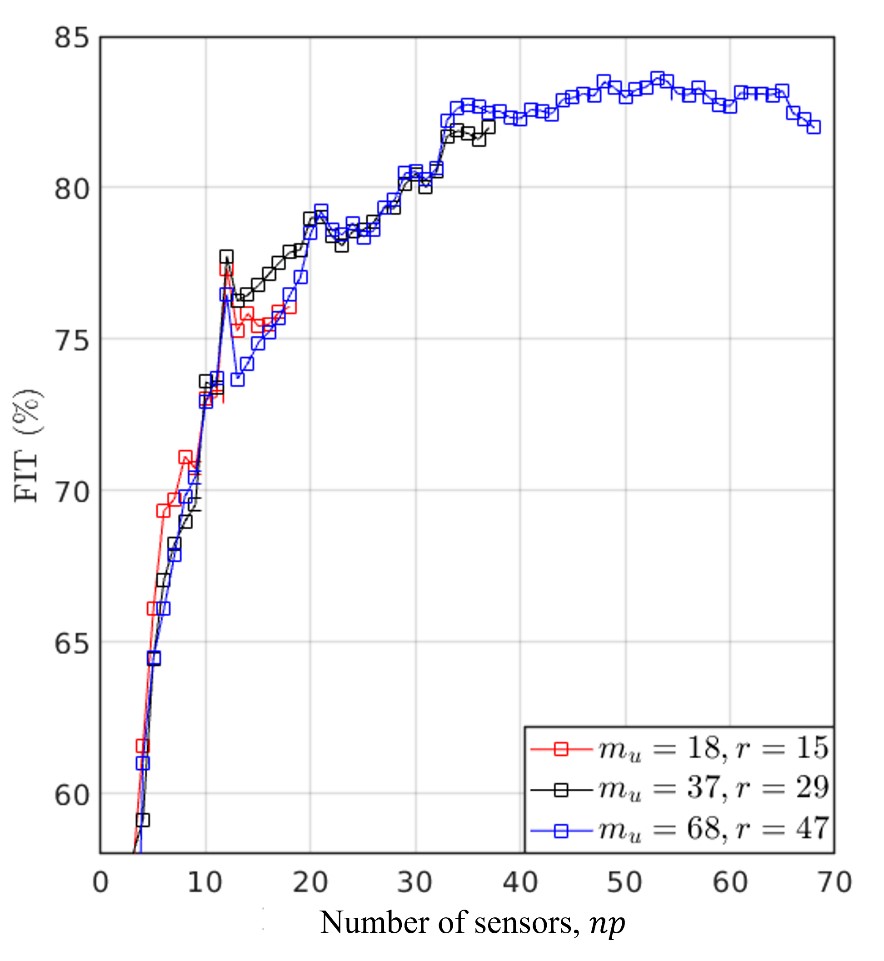}
	}
	\caption{FIT [$\%$] against the number of velocity sensors, $np$.}
	\label{fig:27_UtoUConFITpara}
\end{figure}

In figure \ref{fig:27_UtoUConFITpara}, the FIT[$\%$] metric for different $m_{u}$ and number of sensors $np$ is shown. The parameter $r$ is the model order and is determined so as to capture $97\%$ of the SVD content of the Hankel matrix. It can be seen that the reconstruction performance is practically insensitive to $m_u$; the three curves almost collapse. The number of sensors $np$ is limited by the order $r$ of the model; larger $m_{u}$ increases the model order and enables the use of more sensors. Note the rapid growth of the  FIT[$\%$] metric with $np$; the reconstruction achieves high accuracy with only approximately $15$ sensors. For larger $np$, the performance plateaus. 

\begin{figure}[h!]
	\centerline{
		\includegraphics[width=\linewidth]{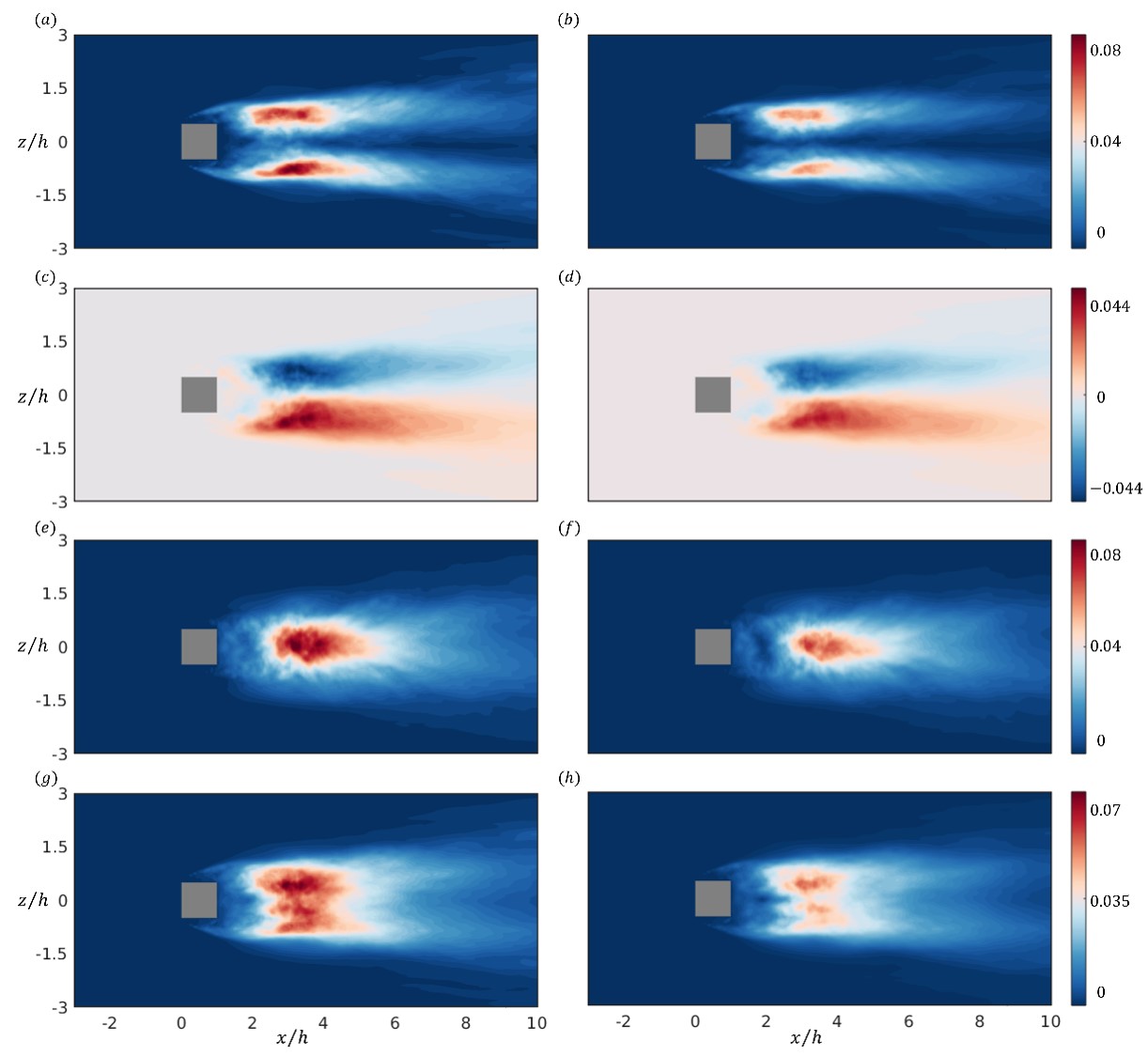}}
	\caption{Flow statistics ($a$-$b$) $\langle u^{\prime 2} \rangle$, ($c$-$d$) $\langle u^{\prime}w^{\prime} \rangle$, ($e$-$f$) $\langle w^{\prime 2} \rangle$, ($g$-$h$) turbulent kinetic energy. Statistics obtained from the true time coefficients $a_i$ (left column), and the estimated time coefficients $\hat{a}_i$ (right column), of the first $10$ POD modes.}
	\label{fig:28_UtoUConRexx}
\end{figure}

The reconstruction quality can be further assessed by comparing the Reynolds stress fields obtained from the true and estimated time coefficients. We select $m_{u}=68, np=53$ that gives the best reconstruction performance. In figure \ref{fig:28_UtoUConRexx}, we compare the Reynolds stress and TKE fields evaluated using the true (left column) and estimated (right column) time coefficients of the first $10$ modes. The same color scale has been employed to facilitate the comparison. It can be seen that the right column reproduces satisfactorily the true flow statistics. The reconstructed normal stresses $\langle u^{\prime}u^{\prime} \rangle$ and $\langle w^{\prime}w^{\prime} \rangle$ have a more confined region of high values compared to the true statistics and this affects also the TKE. The $\langle u^{\prime}w^{\prime} \rangle$ reconstruction is satisfactory.

\begin{figure}[h!]
	\centerline{\includegraphics[width=0.6\linewidth]{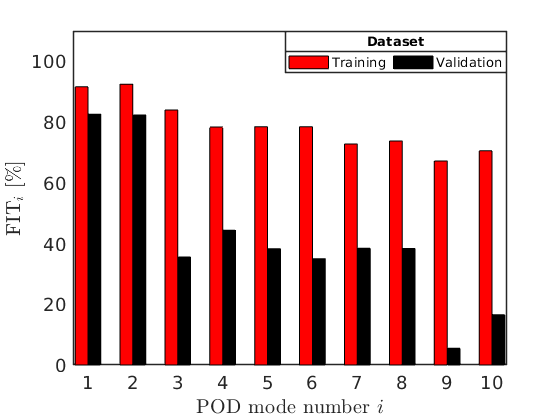}}
	\caption{$\text{FIT}_{i}$[$\%$] of the first $10$ modes for $m_{u}=68, r=38, np=53$. Red: training dataset. Black: validation dataset.}
	\label{fig:29_UtoUConFITibar}
\end{figure}

To explain the observed behavior, the metric $\text{FIT}_{i}[\%]$ for each individual mode, defined as 
\begin{equation}
	\textup{FIT}_i\: [\%] = 100 \left (1 -  \frac{ \left \| a_i[k] -  \hat{a}_i[k] \right \|}{\left \| a_i[k] -   \overline{a_i[k]} \right \|} \right ),
	\label{eq:FITi of each mode}
\end{equation}
was evaluated for the training and validation datasets; the results are shown in figure \ref{fig:29_UtoUConFITibar}.  The reconstruction quality in the training dataset is remarkably good for all the modes considered here. It can be seen that the first two modes are very well reconstructed in the validation dataset, the $\text{FIT}_{i}[\%]$ is more than 80\%. Modes between $3-8$ have lower reconstruction quality, about $35-40\%$, and modes $9$ and $10$ less than $20\%$. The first two modes account for $23\%$ of the total energy and make up $67\%$ of the energy for the first $10$ modes. The modest reconstruction quality for higher modes is responsible for the discrepancies in the flow statistics shown in figure \ref{fig:28_UtoUConRexx}. 

To assess the quality of forecasting of the future evolution of the flow field, we now explore  $q=120$, $239$, $478$, $717$, 
$956$, that correspond to time windows $q \times \Delta t=2.28$, $4.54$, $9.08$, $13.62$ and $18.16$ respectively. Note that the time delay $q \times \Delta t=9.08$ is close to the period of the first two POD modes, and is more than one order of magnitude larger than the Lyapunov time scale, see figure \ref{fig:03_XYXZ_Kolomt}. The largest forecasting window considered, $q \times \Delta=18.16$, is two orders of magnitude larger. 

\begin{figure}[h!]
	\centerline{
		\includegraphics[width=0.6\linewidth]{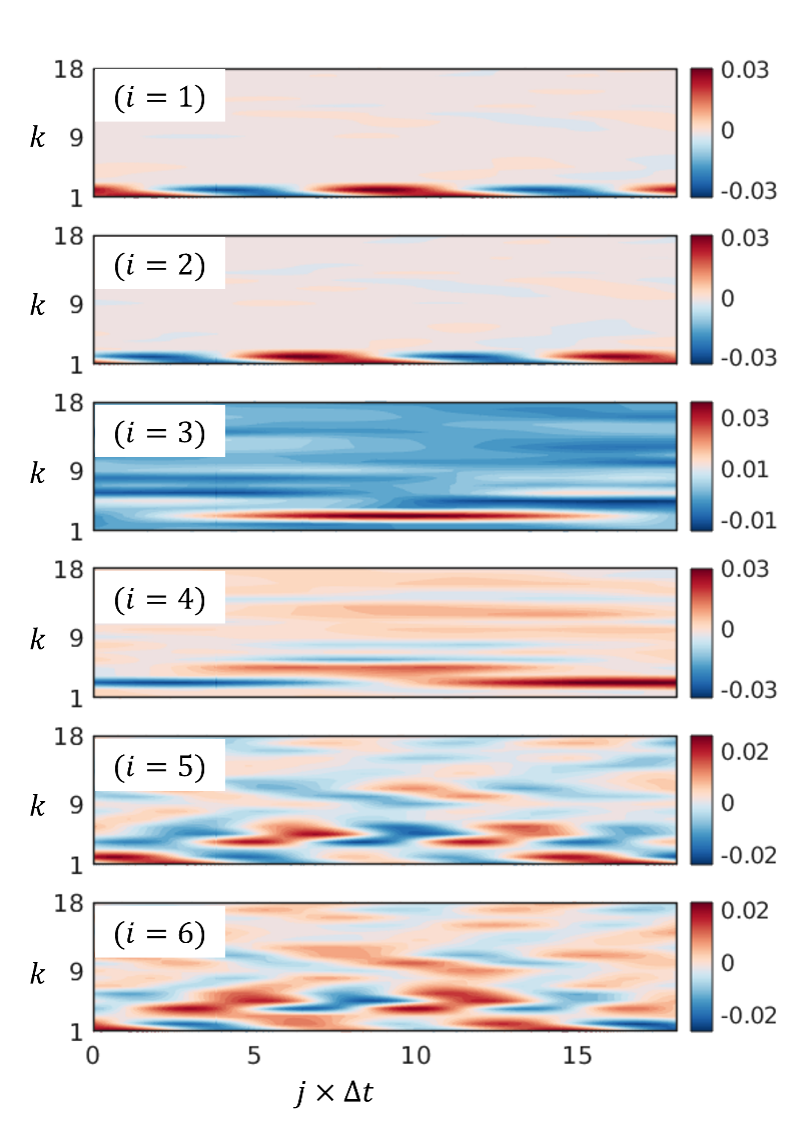}}
	\caption{Contours of the left singular vectors $\boldsymbol{U}^{(u,v,w)}_{H,i}$ in the time-delay / mode order plane, for $m_{u}=18, q=956, q\times \Delta t=18.16$.}
	\label{fig:3rd_UtoUHSVD}
\end{figure}

We first visualise the left singular vectors $\boldsymbol{U}^{(u,v,w)}_{H, i}$ that form the columns of matrix $\boldsymbol{U}_H$, see equation \eqref{eq:U_H_detail}. Contour plots of the first $6$ of those vectors in the time-delay / mode number plane are shown in figure \ref{fig:3rd_UtoUHSVD} for $m_{u}=18$, $q=956$, $q \times \Delta t=18.16$ (approximately equal to 2 shedding periods). It can be seen that the first two singular vectors encode the time-periodic behavior of modes $1$ and $2$; indeed the footprint is strongest for $k=1$ and $k=2$. Note also that the two subplots are shifted in the time-delay axis. In the third singular vector the footprint is strong for $k=3$ over a long time delay, reflecting the features of the slow third POD mode. Higher order singular vectors have more complicated patterns over all values of $k$ and are more difficult to interpret. For three equation Lorenz system \cite{Brunton_et_al_2017} found that the shape the singular vectors take the form of polynomials. This was  because the time delay was short. In the present case, where the time delay is longer, the modes become periodic, as expected from theoretical analysis, see \cite{Frame_Towne_2023}. The message from this plot is clear; the time history of the flow has left its imprint in these vectors and it is exactly this information that allows the forecasting of the future evolution of the flow from current conditions. 

\begin{figure}[h!]
	\centering
	\includegraphics[width=\columnwidth]{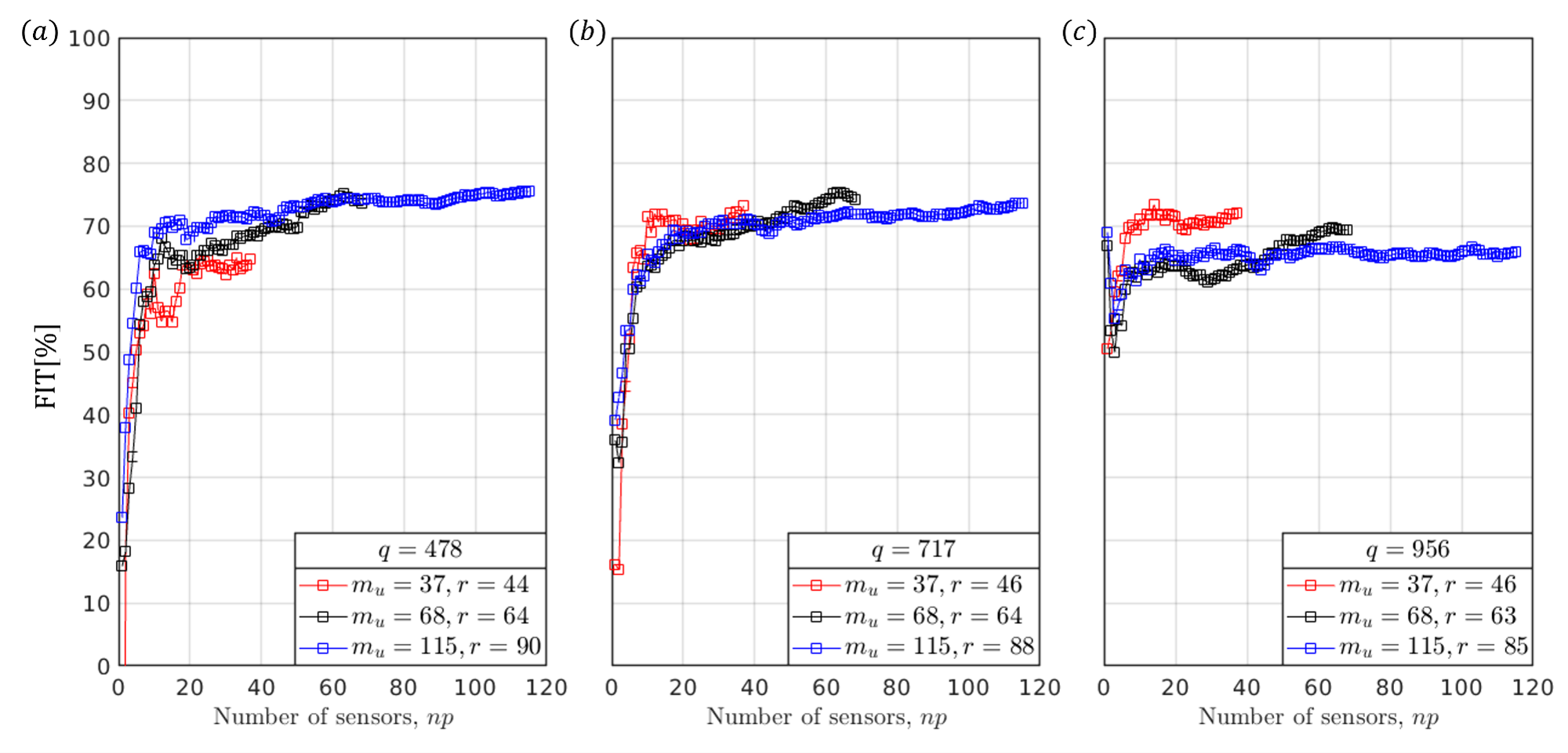}
	\caption{FIT [$\%$] against the number of velocity sensors, $np$: ($a$) $q\times \Delta t=9.08$, ($b$) $q\times \Delta t=13.62$, ($c$) $q\times \Delta t=18.16$.}
	\label{fig:03_UtoUPred_FITpara}
\end{figure}

We first examine how the FIT[$\%$] metric changes with $m_u$ and $np$; results are shown in figure \ref{fig:03_UtoUPred_FITpara} for the 3 largest values of $q$ examined. It can be seen that the FIT[$\%$] curves almost collapse and the forecasting quality is almost independent of $m_{u}$. Only for the largest $q$ there is some small deviation between the 3 curves. Note that again, as with $q=1$, FIT[$\%$] reaches a plateau with relatively few sensors. Most importantly, as $q$ increases the forecasting quality decreases but only slightly; this indicates that it is robust to the size of the foresting window.

\begin{figure}[h!]
	\centering
	\includegraphics[width=\columnwidth]{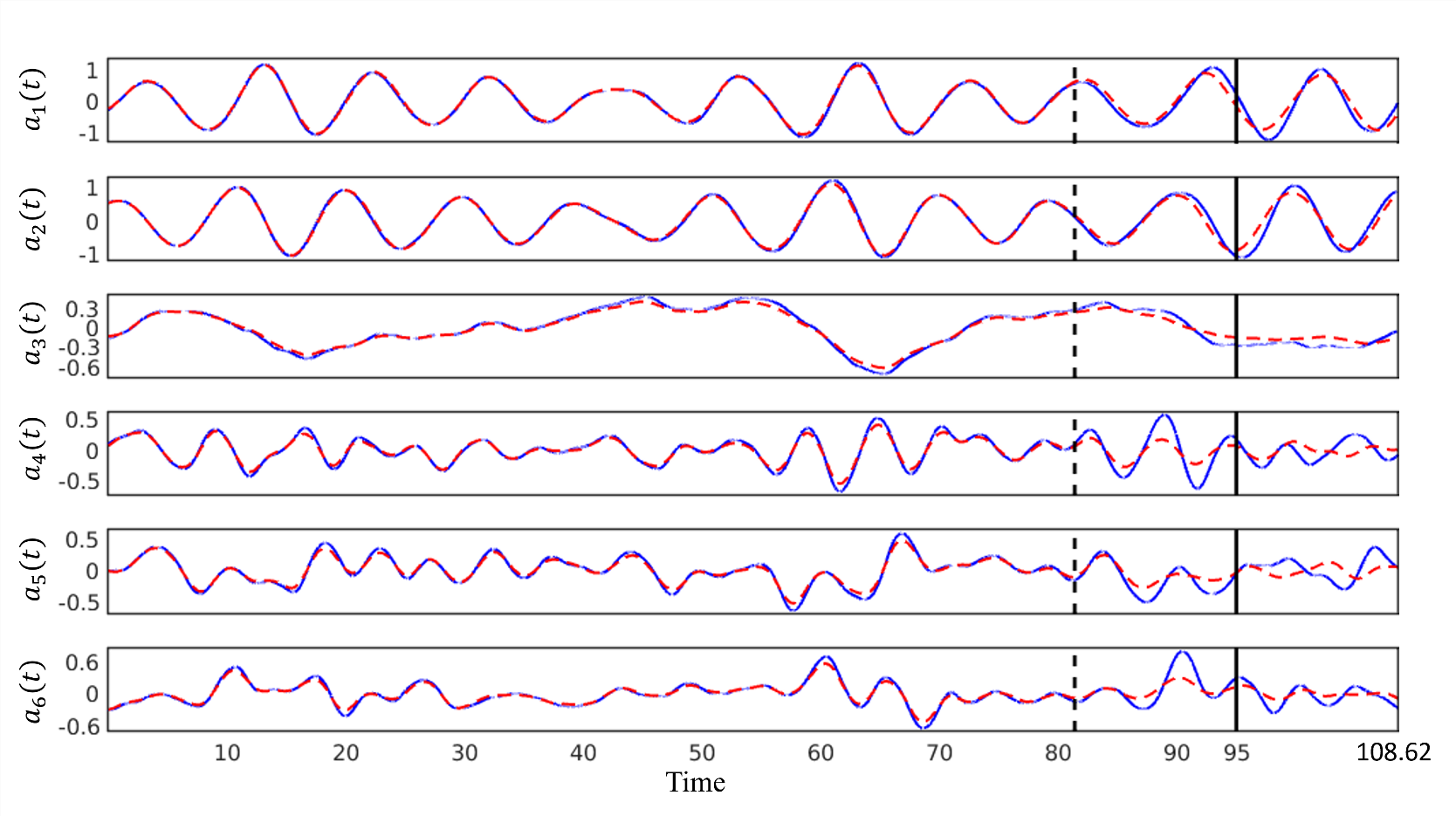}
	\caption{Forecasting of the future evolution of the POD coefficients using velocity measurements for $q\times \Delta t=13.62$ with $m_u=68$, $np=63$. The solid vertical line indicates the starting point of the forecasting dataset. The time-window between the dashed and solid lines marks $t_{p}$ to $t_{Ktrain}$, which is the time extent of the last column of the Hankel matrix $\boldsymbol{H}$. Blue lines indicate DNS and red lines reconstruction/forecasting.}
	\label{fig:03_UtoUPred_ae_q_time_dt=13.62}
\end{figure}

\begin{figure}[h!]
	\centering
	\includegraphics[width=\columnwidth]{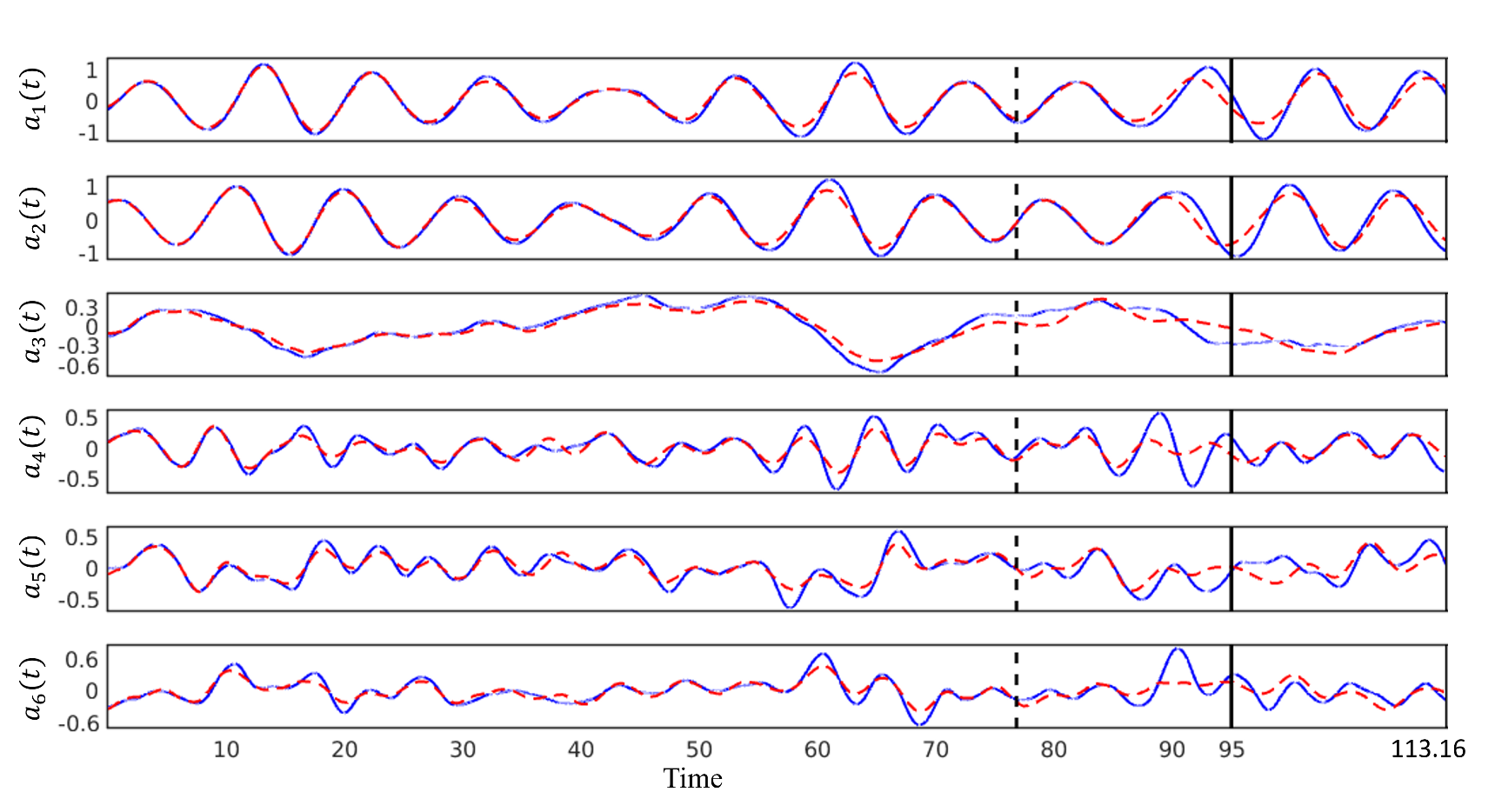}
	\caption{Forecasting of the future evolution of the POD coefficients using velocity measurements for $q\times \Delta t=18.16$ with $m_u=37$, $np=14$. Blue lines indicate DNS and red lines reconstruction/forecasting. For the meaning of the vertical dashed and solid lines, refer to the caption of figure \ref{fig:03_UtoUPred_ae_q_time_dt=13.62}.}
	\label{fig:03_UtoUPred_q_times_dt=18.16}
\end{figure}

This is further demonstrated in figures in \ref{fig:03_UtoUPred_ae_q_time_dt=13.62} and \ref{fig:03_UtoUPred_q_times_dt=18.16} that depict the reconstruction of time coefficients of the 6 dominant POD modes in the training dataset and the future evolution in the forecasting dataset for the two largest time widows $q \times \Delta t=13.62, 18.16$ respectively. The process to obtain these figures is as follows. The training data set (that consists of $K_{train}=5000$ snapshots that correspond to 95 time units) is used to construct the linear model and the estimator as explained in section \ref{sec:forecast_velocity}. Velocity measurements ${\boldsymbol{s}}[k]$ at $np$ points are employed to estimate the time coefficients ${\boldsymbol{a}}[k]$ from $k=0 \dots K_{train}$. At $k=K_{train}$, the estimation is extended from $k_{train+1} \dots K_{train}+q$. Given the 3D turbulent nature of the flow, the forecasting quality of the first two dominant modes is impressive. Both the amplitude and phase are well predicted even for the largest time delay (which is more than two orders of magnitude larger than the Lyapunov time scale). The slowly-evolving third mode is also well forecast. Small discrepancies are detected in the higher modes, but again the phase and amplitude are satisfactorily reproduced especially for the largest time window. Recall that modes $4-6$ have very little energy, between $(1-2)\%$ as shown in figure \ref{fig:19_UVWLambda}, so these deviations are not surprising. Note also that the estimated ${\boldsymbol{a}}[k]$ between $k=K_{train}-q \dots K_{train}$ (region between dashed and solid lines) slightly deviates from the ground truth, probably because this is the last column in the Hankel matrix $\boldsymbol{H}$. These small deviations are more pronounced for modes $4-6$ and also propagate inside the forecasting window affecting the accuracy. 

\subsection{Flow field reconstruction and forecasting from scalar measurements} \label{subsec:ScaOnly}

We now turn our attention to the forecasting of the velocity field from scalar measurements. Recall that we place the scalar sensors at the peaks of the POD velocity modes, but they record only one piece of information, the concentration.

\begin{figure}[h!]
	\centerline{
		\includegraphics[width=\linewidth]{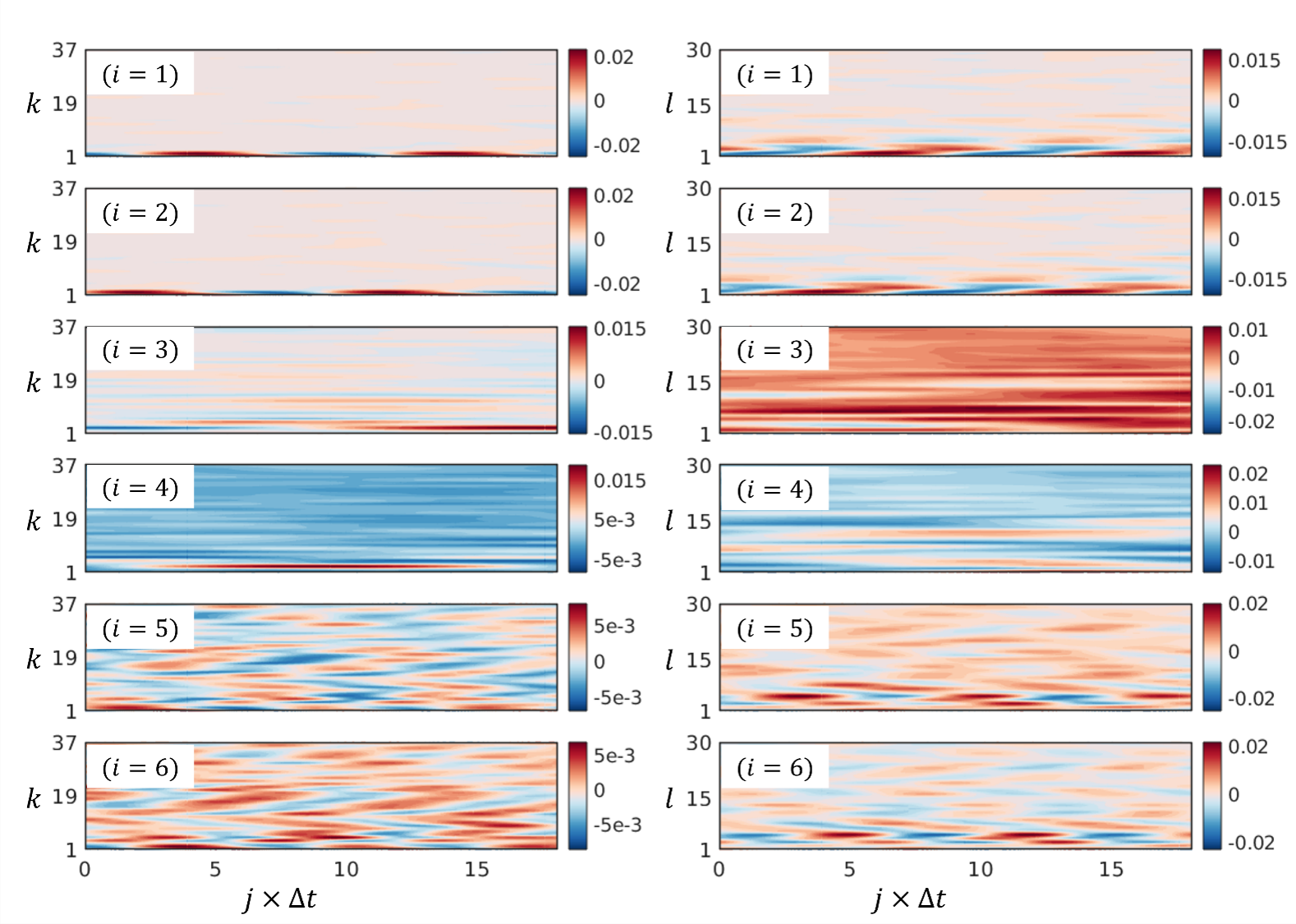}}
	\caption{Contours of the left singular vectors ($a$)  $\boldsymbol{U}^{(u,v,w)}_{H,i}$ and ($b$) $\boldsymbol{U}^{(c)}_{H,i}$ in the time-delay/mode order plane for $m_{u}=37, m_{c}=30, q=956, q\times \Delta t=18.16$.}
	\label{fig:3rd_CtoUHSVD}
\end{figure}

In Fig. \ref{fig:3rd_CtoUHSVD}, we plot contours of the left singular vectors $\boldsymbol{U}^{(u,v,w)}_{H,i} \in \mathbb{R}^{m_{u}}$ and $\boldsymbol{U}^{(c)}_{H,j} \in \mathbb{R}^{m_{c}}$ in the time delay/mode order plane for $m_{u}=37$, $m_{c}=30$, $q=956$, $q \times \Delta t=18.16$.
It can be seen that the periodic behavior of the first two velocity is also detected; the first two scalar modes are also periodic, in agreement with theory. 

\begin{figure}[h!]
	\centering
	\includegraphics[width=\columnwidth]{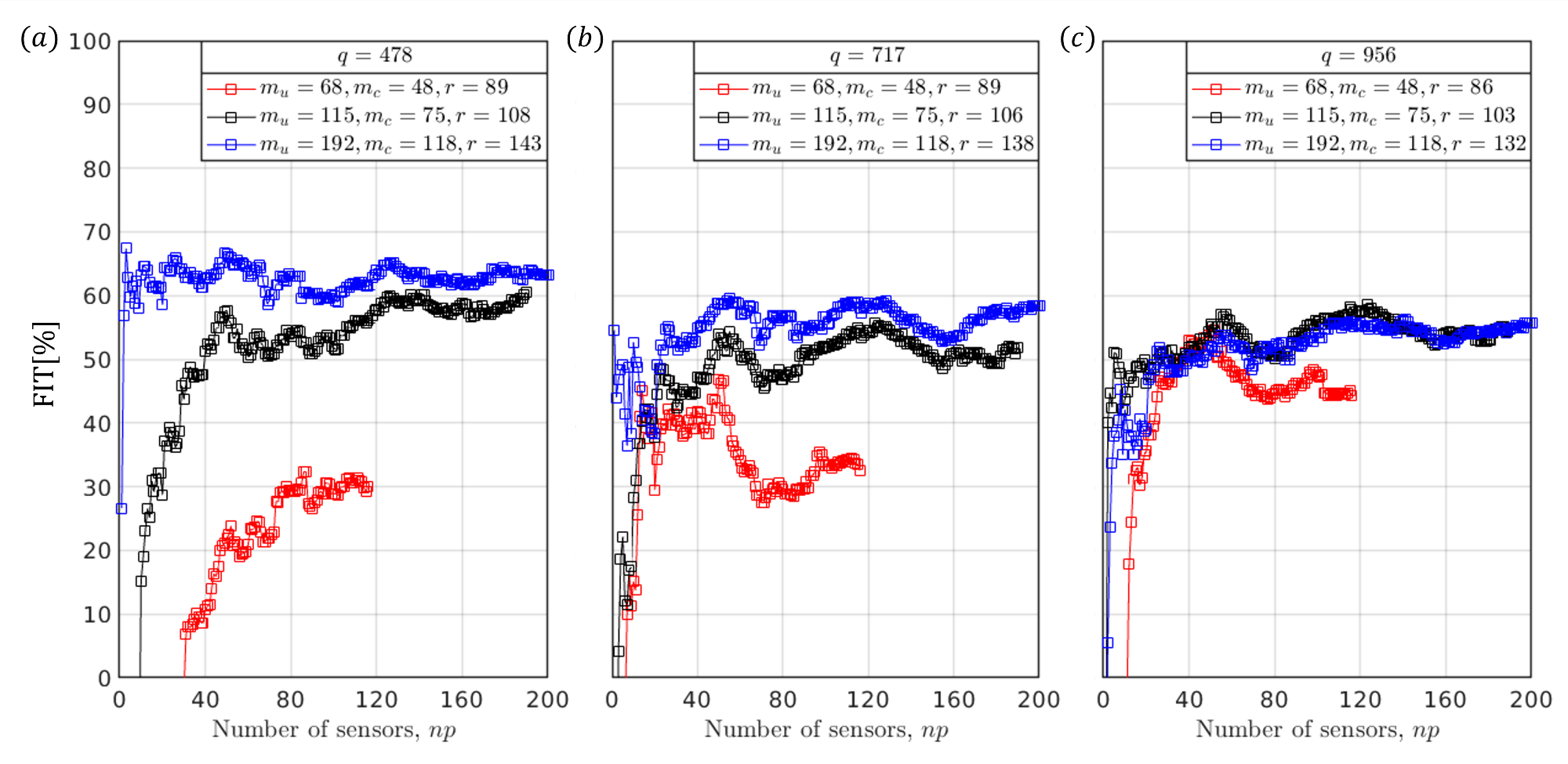}
	\caption{FIT [$\%$] against the number of scalar sensors, $np$ ($a$) $q\times \Delta t=9.08$, ($b$) $q\times \Delta t=13.62$, ($c$) $q\times \Delta t=18.16$.}
	\label{fig:03_CtoUPred_FITpara}
\end{figure}

In Fig. \ref{fig:03_CtoUPred_FITpara}, we explore the effect of $m_c$, $m_c$ and the number of measurements $np$ on the FIT[$\%$] metric for the same values of $q$ as in figure \ref{fig:03_UtoUPred_FITpara}. It can be seen that now larger values of $m_u$ and $m_c$ are required for the results to converge. As in the case of velocity measurements, the  FIT[$\%$] is reduced, but only slightly, as $q$ increases. Again this is a positive result that demonstrates that one can robustly forecast the velocity field from scalar measurements which are more cost effective to obtain. The FIT[$\%$] values are slightly smaller compared to the results of the previous section. Most likely this is because the sensors record only the scalar concentration, i.e.\ they provide less information to the estimator. 

\begin{figure}[h!]
	\centering
	\includegraphics[width=\columnwidth]{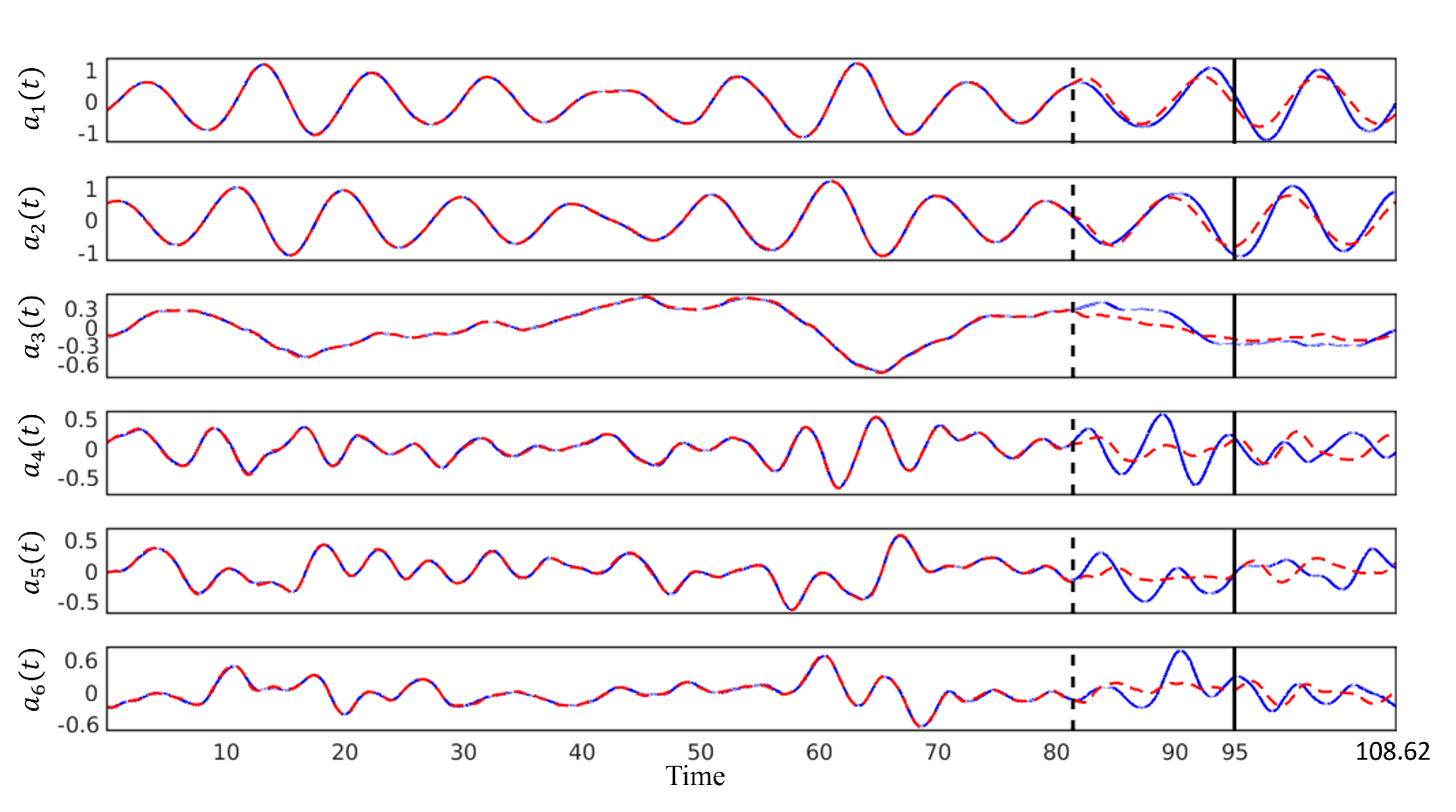}
	\caption{Forecasting of the future evolution of the dominant velocity POD coefficients using scalar measurements for $q\times \Delta t=13.62$ with $m_u=192$, $m_c=118$ and $np=55$.}
	\label{fig:03_CtoUPred_ae_q_times_dt=13.62}
\end{figure}

\begin{figure}[h!]
	\centering
	\includegraphics[width=\columnwidth]{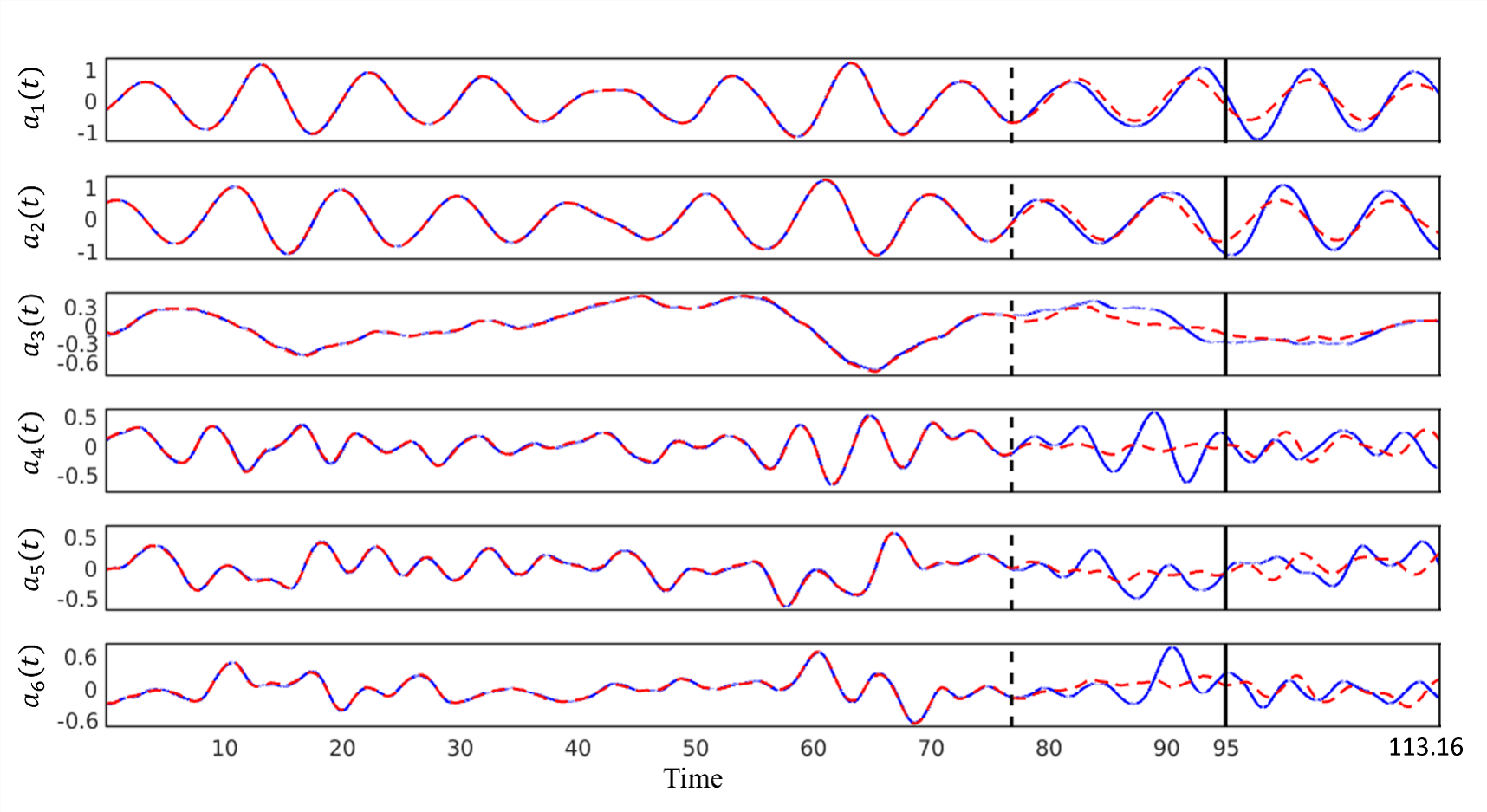}
	\caption{Forecasting of the future evolution of the dominant POD coefficients using scalar measurements for $q\times \Delta t=18.16$ with $m_u=115$, $m_c=75$ and $np=123$.}
	\label{fig:03_CtoUPred_ae_q_times_dt=18.16}
\end{figure}

In figures \ref{fig:03_CtoUPred_ae_q_times_dt=13.62} and \ref{fig:03_CtoUPred_ae_q_times_dt=18.16} we plot the reconstructed and forecast time series of the first 6 POD modes from scalar measurements for the same time windows as in figures \ref{fig:03_UtoUPred_ae_q_time_dt=13.62} and \ref{fig:03_UtoUPred_q_times_dt=18.16} respectively. Again, the time signals closely follow the true ones in the training dataset. In the forecasting dataset, the model predicts the first three time coefficients very well. The quality is comparable to that of figures \ref{fig:03_UtoUPred_ae_q_time_dt=13.62} and  \ref{fig:03_UtoUPred_q_times_dt=18.16} where velocity sensors are used. These plots demonstrate that it is possible to get accurate results from scalar measurements.

\section{Conclusions}\label{sec:conclusions}
A data-driven estimator was synthesized that can forecast the future evolution of a 3D turbulent flow field from current sparse velocity and/or scalar measurements. This is made possible by combining time-delayed embedding, Koopman theory and optimal estimation theory. The key idea is the construction of a linear dynamical system that governs the future POD coefficients and closure of the system using sensor measurements. 

The estimator was applied to the 3D turbulent recirculating flow around a surface mounted cube. Velocity (and scalar, if required) sensors were placed at the peaks of the velocity POD modes. Forecasting the future evolution of the dominant POD modes was performed over time windows one to two orders of magnitude larger than the Lyapunov time scale. Accurate forecasting was obtained for the first three velocity POD modes. Results with velocity sensors were slightly more accurate compared to scalar sensors. The forecasting accuracy only slightly decreased as the window size increased. This was true for both velocity and scalar sensors. The results demonstrate that long forecasts can be made for the most dominant structures of the flow. 

The method can be extended to include other dimensionality reduction techniques, for example convolutional autoencoders. This will allow more compact representation of the dynamics, which is especially useful for high-Reynolds number flows. The resulting nonlinearity of the mapping between the latent space variables and the velocity field can be dealt with other estimators, such as the ensemble Kalman filter. Application of the method to other flow settings, such as wall-bounded flows using pressure and/or shear stress measurements, would be particularly interesting. The framework can also easily incorporate data from moving sensors, such as drones or wearable sensors. Another extension is to train at different operating conditions (for example different Re members) and forecast the flow at an unseen condition using only streaming data from the new condition.

    \bibliographystyle{jfm}


\begin{thebibliography}{48}
		\providecommand{\natexlab}[1]{#1}
		\providecommand{\url}[1]{\texttt{#1}}
		\expandafter\ifx\csname urlstyle\endcsname\relax
		\providecommand{\doi}[1]{doi: #1}\else
		\providecommand{\doi}{doi: \begingroup \urlstyle{rm}\Url}\fi
		
		\bibitem[Brunton and Noack(2015)]{Brunton_Noack_2015}
		S.L. Brunton and B.R. Noack.
		\newblock {Closed-Loop Turbulence Control: Progress and Challenges}.
		\newblock \emph{Applied Mechanics Reviews}, 67\penalty0 (5), 08 2015.
		\newblock 050801.
		
		\bibitem[Sipp and Schmid(2016)]{Sipp_Schmid_2016}
		D.~Sipp and P.J. Schmid.
		\newblock Linear closed-loop control of fluid instabilities and noise-induced
		perturbations: A review of approaches and tools.
		\newblock \emph{Applied Mechanics Reviews}, 68\penalty0 (2), 05 2016.
		\newblock 020801.
		
		\bibitem[Callaham et~al.(2019)Callaham, Maeda, and
		Brunton]{Callaham_et_al_2019}
		J.L. Callaham, K.~Maeda, and S.L. Brunton.
		\newblock Robust flow reconstruction from limited measurements via sparse
		representation.
		\newblock \emph{Phys. Rev. Fluids}, 4:\penalty0 103907, Oct 2019.
		
		\bibitem[Karniadakis et~al.(2021)Karniadakis, Kevrekidis, Lu, Perdikaris, Wang,
		and Yang]{Karniadakis_et_al_2021}
		George~Em Karniadakis, Ioannis~G. Kevrekidis, Lu~Lu, Paris Perdikaris, Sifan
		Wang, and Liu Yang.
		\newblock Physics-informed machine learning.
		\newblock \emph{Nature Reviews Physics}, 3:\penalty0 422–440, 2021.
		
		\bibitem[Box et~al.(2015)Box, Jenkins, Reinsel, and
		Ljung]{Box_Jenkins_Reinsel_Ljung_2015}
		George E.~P. Box, Gwilym~M. Jenkins, Gregory~C. Reinsel, and Greta~M. Ljung.
		\newblock \emph{Time Series Analysis: Forecasting and Control,}.
		\newblock Wiley, fifth edition, 2015.
		
		\bibitem[Lorenz(1963)]{Lorenz1963DeterministicFlows}
		Edward~N Lorenz.
		\newblock {Deterministic Non-Periodic Flows}.
		\newblock \emph{Journal of the Atmospheric Sciences}, 20:\penalty0 130--141,
		1963.
		
		\bibitem[Ruelle(1979)]{Ruelle_1979}
		David Ruelle.
		\newblock Microscopic fluctuations and turbulence.
		\newblock \emph{Physics Letters A}, 72\penalty0 (2):\penalty0 81--82, 1979.
		\newblock ISSN 0375-9601.
		
		\bibitem[Crisanti et~al.(1993)Crisanti, Jensen, Vulpiani, and
		Paladin]{Crisanti_et_al_1993}
		A.~Crisanti, M.~H. Jensen, A.~Vulpiani, and G.~Paladin.
		\newblock Intermittency and predictability in turbulence.
		\newblock \emph{Phys. Rev. Lett.}, 70:\penalty0 166--169, Jan 1993.
		
		\bibitem[Ge et~al.(2023)Ge, Rolland, and
		Vassilicos]{Ge_Rolland_Vassilicos_2023}
		Jin Ge, Joran Rolland, and John~Christos Vassilicos.
		\newblock The production of uncertainty in three-dimensional {Navier–Stokes}
		turbulence.
		\newblock \emph{Journal of Fluid Mechanics}, 977:\penalty0 A17, 2023.
		
		\bibitem[Mohan et~al.(2017)Mohan, Fitzsimmons, and Moser]{Mohan_et_al_2017}
		Prakash Mohan, Nicholas Fitzsimmons, and Robert~D. Moser.
		\newblock Scaling of {Lyapunov} exponents in homogeneous isotropic turbulence.
		\newblock \emph{Phys. Rev. Fluids}, 2:\penalty0 114606, Nov 2017.
		
		\bibitem[Hassanaly and Raman(2019)]{Hassanaly_2019}
		Malik Hassanaly and Venkat Raman.
		\newblock Lyapunov spectrum of forced homogeneous isotropic turbulent flows.
		\newblock \emph{Physical Review Fluids}, 4\penalty0 (11), November 2019.
		\newblock ISSN 2469-990X.
		
		\bibitem[Eivazi et~al.(2021)Eivazi, Guastoni, Schlatter, Azizpour, and
		Vinuesa]{Eivazi_et_al_2021}
		H.~Eivazi, L.~Guastoni, P.~Schlatter, H.~Azizpour, and R.~Vinuesa.
		\newblock Recurrent neural networks and {Koopman}-based frameworks for temporal
		predictions in a low-order model of turbulence.
		\newblock \emph{International Journal of Heat and Fluid Flow}, 90:\penalty0
		108816, 2021.
		
		\bibitem[Vlachas et~al.(2020)Vlachas, Pathak, Hunt, Sapsis, Girvan, Ott, and
		Koumoutsakos]{Vlachas_et_al_2020}
		P.R. Vlachas, J.~Pathak, B.R. Hunt, T.P. Sapsis, M.~Girvan, E.~Ott, and
		P.~Koumoutsakos.
		\newblock Backpropagation algorithms and reservoir computing in recurrent
		neural networks for the forecasting of complex spatiotemporal dynamics.
		\newblock \emph{Neural Networks}, 126:\penalty0 191--217, 2020.
		\newblock ISSN 0893-6080.
		
		\bibitem[Pathak et~al.(2017)Pathak, Lu, Hunt, Girvan, and
		Ott]{Pathak_et_al_2017}
		Jaideep Pathak, Zhixin Lu, Brian~R. Hunt, Michelle Girvan, and Edward Ott.
		\newblock Using machine learning to replicate chaotic attractors and calculate
		lyapunov exponents from data.
		\newblock \emph{Chaos: An Interdisciplinary Journal of Nonlinear Science},
		27\penalty0 (12):\penalty0 121102, 12 2017.
		\newblock ISSN 1054-1500.
		
		\bibitem[Khodkar and Hassanzadeh(2021)]{Khodkar2021}
		M.~A. Khodkar and P.~Hassanzadeh.
		\newblock A data-driven, physics-informed framework for forecasting the
		spatiotemporal evolution of chaotic dynamics with nonlinearities modeled as
		exogenous forcings.
		\newblock \emph{Journal of Computational Physics}, 440, 2021.
		\newblock ISSN 110412.
		
		\bibitem[Dubois et~al.(2020)Dubois, Gomez, Planckaert, and
		Perret]{Dubois_et_al_2020}
		P.~Dubois, T.~Gomez, L.~Planckaert, and L.~Perret.
		\newblock Data-driven predictions of the {Lorenz} system.
		\newblock \emph{Physica D}, 408:\penalty0 132495, 2020.
		
		\bibitem[Allen et~al.(2025)Allen, Markou, Tebbutt, and
		et~al.]{Allen_et_al_2025}
		A.~Allen, S.~Markou, W.~Tebbutt, and et~al.
		\newblock End-to-end data-driven weather prediction.
		\newblock \emph{Nature}, \penalty0
		(https://doi.org/10.1038/s41586-025-08897-0), 2025.
		
		\bibitem[Zhang et~al.(2019)Zhang, Sun, Magnusson, Buizza, Lin, Chen, and
		Emanuel]{Zhang_Sun_et_al_2019}
		Fuqing Zhang, Y.~Qiang Sun, Linus Magnusson, Roberto Buizza, Shian-Jiann Lin,
		Jan-Huey Chen, and Kerry Emanuel.
		\newblock What is the predictability limit of midlatitude weather?
		\newblock \emph{Journal of the Atmospheric Sciences}, 76\penalty0 (4):\penalty0
		1077 -- 1091, 2019.
		
		\bibitem[Pope(2000)]{pope_2000}
		Stephen~B. Pope.
		\newblock \emph{Turbulent Flows}.
		\newblock Cambridge University Press, 2000.
		
		\bibitem[Moehlis et~al.(2004)Moehlis, Faisst, and Eckhardt]{Moehlis_et_al_2004}
		Jeff Moehlis, Holger Faisst, and Bruno Eckhardt.
		\newblock A low-dimensional model for turbulent shear flows.
		\newblock \emph{New Journal of Physics}, 6\penalty0 (1):\penalty0 56, may 2004.
		
		\bibitem[Schmid(2010)]{Schmid2010JFM}
		P.~J. Schmid.
		\newblock Dynamic mode decomposition of numerical and experimental data.
		\newblock \emph{Journal of Fluid Mechanics}, 656, 2010.
		
		\bibitem[Le~Clainche and Vega(2017)]{LeClainche_Vega_2017}
		Soledad Le~Clainche and Jos\'{e}~M. Vega.
		\newblock Higher order dynamic mode decomposition.
		\newblock \emph{SIAM Journal on Applied Dynamical Systems}, 16\penalty0
		(2):\penalty0 882--925, 2017.
		
		\bibitem[Chu and Schmidt(2025)]{Chu_Schmidt_2025}
		Tianyi Chu and Oliver Schmidt.
		\newblock Stochastic reduced-order {Koopman} model for turbulent flows.
		\newblock 2025.
		\newblock \doi{10.48550/arXiv.2503.22649}.
		
		\bibitem[Brunton et~al.(2017)Brunton, Brunton, Proctor, Kaiser, and
		Kutz]{Brunton_et_al_2017}
		S.~L. Brunton, B.~W. Brunton, J.~L. Proctor, E.~Kaiser, and J.~N. Kutz.
		\newblock Chaos as an intermittently forced linear system.
		\newblock \emph{Nature Communications}, 8\penalty0 (19), 2017.
		
		\bibitem[Dylewsky et~al.(2022)Dylewsky, Kaiser, Brunton, and
		Kutz]{Dylewsky_et_al_2022}
		Daniel Dylewsky, Eurika Kaiser, Steven~L. Brunton, and J.~Nathan Kutz.
		\newblock Principal component trajectories for modeling spectrally continuous
		dynamics as forced linear systems.
		\newblock \emph{Phys. Rev. E}, 105:\penalty0 015312, Jan 2022.
		
		\bibitem[Takens(1981)]{Takens_1981}
		Floris Takens.
		\newblock Detecting strange attractors in turbulence.
		\newblock In David Rand and Lai-Sang Young, editors, \emph{Dynamical Systems
			and Turbulence, Warwick 1980}, pages 366--381. Springer:Berlin, Heidelberg,
		1981.
		\newblock ISBN 978-3-540-38945-3.
		
		\bibitem[Kamb et~al.(2020)Kamb, Kaiser, Brunton, and Kutz]{Kamb_et_al_2020}
		Mason Kamb, Eurika Kaiser, Steven~L. Brunton, and J.~Nathan Kutz.
		\newblock Time-delay observables for {Koopman}: Theory and applications.
		\newblock \emph{SIAM Journal on Applied Dynamical Systems}, 19\penalty0
		(2):\penalty0 886--917, 2020.
		
		\bibitem[Arbabi and Mezi\'c(2017)]{Arbabi2017}
		H.~Arbabi and I.~Mezi\'c.
		\newblock Ergodic theory, dynamic mode decomposition, and computation of
		spectral properties of the koopman operator.
		\newblock \emph{SIAM J. Applied Dynamical Systems}, 16\penalty0 (4), 2017.
		
		\bibitem[Pan and Duraisamy(2019)]{Pan_Duraisamy_2019}
		Shaowu Pan and Karthik Duraisamy.
		\newblock On the structure of time-delay embedding in linear models of
		non-linear dynamical systems.
		\newblock \emph{Chaos}, 30 7:\penalty0 073135, 2019.
		
		\bibitem[Pan and Duraisamy(2020)]{Pan_Duraisamy_2020}
		Shaowu Pan and Karthik Duraisamy.
		\newblock Physics-informed probabilistic learning of linear embeddings of
		nonlinear dynamics with guaranteed stability.
		\newblock \emph{SIAM Journal on Applied Dynamical Systems}, 19\penalty0
		(1):\penalty0 480--509, 2020.
		
		\bibitem[Giannakis(2019)]{Giannakis_2019}
		Dimitrios Giannakis.
		\newblock Data-driven spectral decomposition and forecasting of ergodic
		dynamical systems.
		\newblock \emph{Applied and Computational Harmonic Analysis}, 47\penalty0
		(2):\penalty0 338--396, 2019.
		\newblock ISSN 1063-5203.
		
		\bibitem[Das and Giannakis(2019)]{Das_Giannakis_2019}
		Suddhasattwa Das and Dimitrios Giannakis.
		\newblock Delay-coordinate maps and the spectra of koopman operators.
		\newblock \emph{Journal of Statistical Physics}, 175:\penalty0 1107–1145,
		2019.
		
		\bibitem[Frame and Towne(2023)]{Frame_Towne_2023}
		Peter Frame and Aaron Towne.
		\newblock Space-time {POD} and the {Hankel} matrix.
		\newblock \emph{PLOS ONE}, 18\penalty0 (8):\penalty0 1--31, 08 2023.
		
		\bibitem[Kailath et~al.(2000)Kailath, Hassibi, and
		Sayed]{Kailath_Hassibi_Sayed_2000}
		T.~Kailath, B.~Hassibi, and A.~H. Sayed.
		\newblock \emph{Linear estimation}.
		\newblock Prentice-Hall International, 2000.
		\newblock ISBN 0130224642.
		
		\bibitem[Schmidt(2025)]{Schmidt_2025}
		Oliver~T. Schmidt.
		\newblock Data-driven forecasting of high-dimensional transient and stationary
		processes via space-time projection.
		\newblock 2025.
		\newblock URL \url{https://arxiv.org/abs/2503.23686}.
		
		\bibitem[Sirovich(1987)]{Sirovich1987}
		L.~Sirovich.
		\newblock Turbulence and the dynamics of coherent structures.
		\newblock \emph{Q. Appl. Maths}, 45:\penalty0 561--571, 1987.
		
		\bibitem[Brunton et~al.(2020)Brunton, Noack, and
		Koumoutsakos]{Brunton_et_al_2020}
		Steven~L. Brunton, Bernd~R. Noack, and Petros Koumoutsakos.
		\newblock Machine learning for fluid mechanics.
		\newblock \emph{Annual Review of Fluid Mechanics}, 52:\penalty0 477--508, 2020.
		\newblock ISSN 1545-4479.
		
		\bibitem[Evensen(2003)]{Evensen_2003}
		Geir Evensen.
		\newblock The {Ensemble Kalman Filter}: theoretical formulation and practical
		implementation.
		\newblock \emph{Ocean Dynamics}, 53:\penalty0 343–367, 2003.
		
		\bibitem[Krajnovic and Davidson(2002)]{Krajnovic2002}
		S.~Krajnovic and L.~Davidson.
		\newblock Large-eddy simulation of the flow around a bluff body.
		\newblock \emph{AIAA JOURNAL}, 40:\penalty0 927--936, 2002.
		
		\bibitem[Yao and Papadakis(2023)]{Yao_papadakis_2023}
		H.~Yao and G.~Papadakis.
		\newblock On the role of the laminar/turbulent interface in energy transfer
		between scales in bypass transition.
		\newblock \emph{Journal of Fluid Mechanics}, 960:\penalty0 A24, 2023.
		
		\bibitem[Schlander et~al.(2024)Schlander, Rigopoulos, and
		Papadakis]{Schlander_Rigopoulos_Papadakis_2024}
		Rasmus~Korslund Schlander, Stelios Rigopoulos, and George Papadakis.
		\newblock Resolvent analysis of turbulent flow laden with low-inertia
		particles.
		\newblock \emph{Journal of Fluid Mechanics}, 985:\penalty0 A27, 2024.
		
		\bibitem[Thomareis and Papadakis(2017)]{Thomareis_Papadakis_2017}
		Nikitas Thomareis and George Papadakis.
		\newblock Effect of trailing edge shape on the separated flow characteristics
		around an airfoil at low {Reynolds} number: A numerical study.
		\newblock \emph{Physics of Fluids}, 29\penalty0 (1):\penalty0 014101, 2017.
		
		\bibitem[Thomareis and Papadakis(2018)]{Thomareis_Papadakis_2018}
		N.~Thomareis and G.~Papadakis.
		\newblock Resolvent analysis of separated and attached flows around an airfoil
		at transitional {Reynolds} number.
		\newblock \emph{Phys. Rev. Fluids}, 3:\penalty0 073901, Jul 2018.
		
		\bibitem[Castro and Robins(1977)]{Castro1977UniFlow}
		I.~P. Castro and A.~G. Robins.
		\newblock The flow around a surface-mounted cube in uniform and turbulent
		streams.
		\newblock \emph{Journal of Fluid Mechanics}, 79:\penalty0 307--335, 1977.
		
		\bibitem[Rossi et~al.(2010)Rossi, Philips, and Iaccarino]{Rossi-2010}
		R.~Rossi, D.~A. Philips, and G.~Iaccarino.
		\newblock A numerical study of scalar dispersion downstream of a wall-mounted
		cube using direct simulations and algebraic flux models.
		\newblock \emph{International Journal of Heat and Fluid Flow}, 31:\penalty0
		805--819, 2010.
		
		\bibitem[Li et~al.(2021)Li, Hulshoff, and Hickel]{Li2021}
		X.~Li, S.~Hulshoff, and S.~Hickel.
		\newblock Towards adjoint-based mesh refinement for large eddy simulation using
		reduced-order primal solutions: Preliminary 1d burgers study.
		\newblock \emph{Comput. Methods Appl. Mech. Engrg}, 379, 2021.
		\newblock ISSN 113733.
		
		\bibitem[Knight and Sirovich(1990)]{knight1990}
		Bruce Knight and Lawrence Sirovich.
		\newblock Kolmogorov inertial range for inhomogeneous turbulent flows.
		\newblock \emph{Physical review letters}, 65\penalty0 (11):\penalty0 1356,
		1990.
		
		\bibitem[Bourgeois et~al.(2011)Bourgeois, Sattari, and
		Martinuzzi]{Bourgeois_et_al_2011}
		J.~A. Bourgeois, P.~Sattari, and R.~J. Martinuzzi.
		\newblock Alternating half-loop shedding in the turbulent wake of a finite
		surface-mounted square cylinder with a thin boundary layer.
		\newblock \emph{Physics of Fluid}, 23:\penalty0 095101, 2011.
		
	\end{thebibliography}

\end{document}